\documentclass[%
reprint,
superscriptaddress,
nofootinbib,
amsmath,amssymb,
aps,
prb,
longbibliography
]{revtex4-2}

\usepackage{graphicx}% Include figure files

\usepackage{bm}% bold math
\usepackage{xcolor}

\usepackage[mathlines]{lineno}% Enable numbering of text and display math
\usepackage[colorlinks=true, linkcolor=blue,urlcolor=blue,citecolor=blue]{hyperref}% add hypertext capabilities
\usepackage{cleveref}

\usepackage[ignoreunlbld,norefs,nocites]{refcheck}

\begin{document}

\preprint{APS/123-QED}

\title{Correlation effects in one-dimensional metallic quantum wires under various confinements}% Force line breaks with \\
%\thanks{A footnote to the article title}%

\author{Vidit Gangwar}
\affiliation{%
Computational Quantum Many-Body Physics Lab, Department of Physics, Dr.\ B.\ R.\ Ambedkar National Institute of Technology, Jalandhar, Punjab - 144008, India}%

\author{Vinod Ashokan}
\email{ashokanv@nitj.ac.in}
\affiliation{Computational Quantum Many-Body Physics Lab, Department of Physics, Dr.\ B.\ R.\ Ambedkar National Institute of Technology, Jalandhar, Punjab - 144008, India}
\author{Ankush Girdhar}
\affiliation{%
Department of Physics, School of Advanced Sciences, Vellore Institute of Technology, Vellore, Tamil Nadu - 632014, India}%

\author{Klaus Morawetz}
\affiliation{M\"{u}nster University of Applied Sciences, Stegerwaldstrasse 39, 48565 Steinfurt, Germany}%

\affiliation{International Institute of Physics - UFRN, Campus Universit\'ario Lagoa nova, 59078-970 Natal, Brazil}%
\author{N.\ D.\ Drummond}
\affiliation{Department of Physics, Lancaster University, Lancaster LA1 4YB, United Kingdom}%

\author{K.\ N.\ Pathak}%
\affiliation{Centre for Advanced Study in Physics, Panjab
 University, Chandigarh - 160014, India}%

\date{\today}% It is always \today, today,
\begin{abstract}

Dynamical response theory is employed to investigate the effects of various transverse confinements on electron correlations in the ground state of a ferromagnetic one-dimensional quantum wire for different wire widths $b$ and density parameters $r_\text{s}$. In the regime of a thin quantum wire, electrons are treated as a one-dimensional gas under different confinement models via effective electron-electron interaction potentials. Using the first-order random phase approximation (FRPA) including self-energy and exchange contribution, which provides the ground state structure beyond the random phase approximation, we numerically compute the structure factor, pair-correlation function, correlation energy, and ground-state energy for various values of $b$ and $r_\text{s}$. Our results reveal that the correlation energy depends on the choice of confinement model. For the ultrathin wire $(b\rightarrow 0)$ in the high-density limit, we find that the correlation energy for transverse confinement models $V_1(q)$ (harmonic), $V_2(q)$ (cylindrical), and $V_5(q)$ (harmonic-delta) approaches $\epsilon_\text{c}(r_\text{s})= - \pi^2/360 \sim -0.02741$ a.u., which agrees with the exact results in this limit [P.-F.\ Loos, J.\ Chem.\
Phys.\ \textbf{138}, 064108 (2013), V.\ Ashokan \textit{et al.}, Phys.\ Rev.\ B \textbf{101}, 075130 (2020)]. This clearly illustrates that for at least these three confinement potentials, the one-dimensional Coulomb potential can be regularized at interparticle distance $x=0$ to yield the same correlation energy. In contrast, other confinement potentials, $V_3(q)$ (infinite square well), $V_4(q)$ (infinite square-infinite triangular well), and $V_6(q)$ (infinite square-delta well), do not approach the same high-density limit; instead, the correlation energy tends to $\epsilon_\text{c} \sim -0.03002$ a.u.\ for these potentials. The percentage difference in correlation energy between the confinement models $V_1(q)$, $V_2(q)$, $V_5(q)$ and $V_3(q)$, $V_4(q)$, $V_6(q)$ is within about $10\%$  in the high-density limit. The ground-state properties obtained from the FRPA are compared with the available quantum Monte Carlo results in the high-density regime. We observe that the peak height in the static structure factor at $k=2k_\text{F}$ depends significantly on the confinement model. These prominent peaks at $k=2k_\text{F}$ are fitted with a function based on our finite wire-width theory, guided by insights from bosonization, demonstrating good agreement with our FRPA theory.  

\end{abstract}

%\keywords{Suggested keywords}%Use showkeys class option if keyword
                              %display desired
 \maketitle

%\tableofcontents

\section{Introduction}
The study of interacting particles has been essential for understanding the complex quantum behaviors observed in condensed matter physics \cite{Giuliani2008,morawetz2024electronic,morawetz2024off,zhao2025universal}. Recent experimental breakthroughs have produced exceptionally clean quasi-one-dimensional (1D) systems, including strongly interacting 1D Fermi gases in optical lattices \cite{moritz2005confinement}, carbon nanotubes \cite{bockrath1999luttinger,ishii2003direct,shiraishi2003tomonaga}, semiconductor nanowires \cite{levy2006luttinger,schafer2008new,dudy2017one,hong2016one}, and high-mobility quantum wires grown by cleaved edge overgrowth \cite{yacoby1996nonuniversal}. These recent developments have generated considerable theoretical interest and research.

In many-body physics \cite{giamarchi2004quantum,Giuliani2008}, theoretical investigations predict qualitatively distinct quantum behaviors in systems of reduced dimensionality (e.g., two dimensions, 1D) as compared to their three-dimensional counterparts. These deviations are anticipated both at zero temperature and at finite temperatures within the thermodynamic limit \cite{ashokan2018one,garg2018finite,ashokan2020exact,rani2024some}.
Reducing the dimensionality of electron systems to one leads to a breakdown of the Landau's Fermi-liquid picture. In this regime, strong Coulomb interactions drive electrons to behave collectively, a phenomenon explained by the Tomonaga-Luttinger liquid theory \cite{tomonaga1950remarks,luttinger1963exactly}, replacing the Landau picture of individual particle description. One-dimensional interacting systems exhibit intriguing phenomena, including spin-charge separation \cite{girdhar2023wire,auslaender2005spin,Ruwan2022}, where collective spin and charge excitations become distinct and propagate independently; charge fractionalization \cite{steinberg2008charge,ramos2024nonlinear}, where charge excitations carry nonquantized charge; and Wigner crystallization \cite{schulz1993wigner,meyer2008wigner,deshpande2008one,girdhar2023wigner}, where electrons get localized due to repulsive Coulomb interactions between them. 

The electronic properties of quasi-1D systems are significantly influenced by electron-electron interactions under various transverse confinement schemes. The effective electron-electron interaction will be repulsive for all energies as long as the uniformly positively charged background is rigid. However, confinement effects manifest themselves as modifications to the electronic band structure, enhancement of Coulomb interactions between electrons, and the emergence of novel collective excitations. 
\par In this work, to examine the role of the nature of confinement and the correlation effects, we consider some different confinement models developed in the literature. Among them, the harmonic confinement model is quite realistic in interparticle potential \cite{Friesen1980}, and is capable of tackling excitonic instability for a rigid and uniformly charged positive background. Interactions by a softened Coulomb potential appropriate for a cylindrical wire \cite{ashokan2020exact,sharma2021ground} also remove the singularity at zero interparticle distance. The interactions can also be modelled using a square quantum well confinement approach when the subband separation is large, i.e., for thin wires, such as in GaAs semiconductor quantum-well wires \cite{Campos1995}. Epitaxial growth techniques are commonly used to fabricate these wires, and their optical properties have been extensively reported \cite{wang2006epitaxial}. In 1D quantum wires within GaAs heterostructures, interactions are studied using a square quantum well confinement model and realistic variational optimized wave functions for the transverse directions \cite{Das1985}. Therefore, an accurate description of the ground-state properties of 1D wires with realistic Coulomb interactions remains a complex challenge, even though significant progress has been made towards more sophisticated and accurate modelling. Moudgil \textit{et al.}\ \cite{Moudgil2010} studied the plasmon dispersion with different transverse confinement models on an atom-scale metallic wire using the random phase approximation (RPA) and the Singwi-Tosi-Land-Sj\"{o}lander (STLS) approximation, which shows good agreement with electron-energy-loss spectroscopy measurements \cite{nagao2006one}. However, the structure factor and correlation energy were not calculated for the different confinement models on the effective electron-electron interaction potential. In this paper, we model 1D homogeneous electron gases (HEGs) in the high density limit interacting via effective long-range Coulomb potentials and report the numerical calculation of the static structure factor (SSF), pair-correlation function (PCF), correlation energy, and ground state energy for various confinement models \cite{Friesen1980, ashokan2020exact, Campos1995, Das1985, Hu1990, Li1989} and compare with the available quantum Monte Carlo (QMC) simulation results including variational Monte Carlo (VMC) and diffusion Monte Carlo (DMC) \cite{ashokan2018one, girdhar2022electron}. 
\par The rest of the paper is structured as follows. In Sec.\ \ref{Theoretical formalism}, we present various theoretical models for the confinement of electrons in a quasi-one-dimensional HEG\@. We also discuss the dynamic density response function using first-order random phase approximation (FRPA)\@. In Sec.\ \ref{SSF}, the SSF is calculated numerically and compared under various confinement schemes for several wire widths $(b)$ and density parameters $(r_\text{s})$. In Sec.\ \ref{PCF}, the PCF is calculated numerically and compared under various confinement schemes for several wire widths and density parameters. In Sec.\ \ref{Groundstateenergy}, we report the ground state and correlation energies' dependence on the wire widths ($b$) and density parameters ($r_\text{s}$). In Sec.\ \ref{results}, we present results and a detailed discussion of other ground-state properties of the 1D HEG for various confinement models at different coupling parameters ($b$ and $r_\text{s}$). Finally, we summarize our work in Sec.\ \ref{Conclusion}. Throughout the paper, we use Hartree atomic units ($\hbar=|e|=m_\text{e}=4\pi\epsilon=1$).
% ----------------------------------------------------------
\section{Theoretical formalism}\label{Theoretical formalism}

\subsection{Quasi-1D wires}

Our model considers a quasi-1D system, a finite-width quantum wire in which electrons can freely roam along the wire axis, neutralized by a rigid positive background and strongly confined in the transverse directions using various confinement potentials. The wire is modelled such that the length of the wire is much larger than its width, allowing electrons to move freely along the $x$-axis (wire axis) and confining them in the $yz$- plane by a potential $V(y,z)$ uniform in $x$. The strength of the transverse confinement can be used to tune the electron-electron interaction in quantum wires. 
Numerical or approximate methods are essential for studying the ground state properties of quasi-1D electron systems under the various transverse confinement models, as exact analytical solutions are usually not possible. We limit our calculations to the extreme quantum limit, a regime in which electrons occupy only the lowest energy subband associated with transverse motion, as this approximation is valid for atomic-scale metallic wires with significant subband separation \cite{Moudgil2010, Campos1995}.
%------------------------------------------------------ 
\subsection{Confinement models}\label{models}
\subsubsection{\textbf{Harmonic confinement in the \textit{yz}-plane and particles free to move along the \textit{x}-direction}}
In this confinement scheme \cite{Friesen1980, Lee2011}, the effective interaction between electrons in a quasi-1D wire is considered by taking the harmonic confinement potential along the transverse directions ($yz$- plane), which confines the electrons tightly and keeps them close to the center of the wire along the $x$-axis. This model represents a cylindrically symmetric chain as a collection of electrons in gas against a rigid positive background. The analytic form of this effective interaction potential in real space is
\begin{equation}
    v_1(x)=\frac{\sqrt{\pi}}{2 b}{\rm e}^\frac{x^2}{ 4b^2} {\rm erfc}\left(\frac{|x|}{ 2 b}\right)
\end{equation}
and the Fourier transform is given as
\begin{equation}\label{V1}
    V_1(q)= \exp \left(q^{2} b^{2}\right) E_{1}\left(q^{2} b^{2}\right),
\end{equation}
where $E_{1}$ denotes the exponential integral function of the first kind. 
The analytic expression is derived in Appendix \ref{AppendixA} for ready reference.

\subsubsection{\textbf{Cylindrical confinement in the \textit{yz}-plane}}
This model represents the effective interactions between cylindrically confined electrons through a softened Coulomb potential \cite{ashokan2020exact} defined as $v_2(x)=1/\sqrt{x^2+b^2}$,
where $b$ is wire width. The Fourier transform is given as
\begin{equation}\label{V2}
    V_2(q)=2K_0(bq),
\end{equation}
where $K_{0}(b q)$ denotes the zeroth-order modified Bessel function of the second kind. The analytic expression is derived in Appendix \ref{AppendixB}.

\subsubsection{\textbf{Infinite square well confinement in the \textit{yz}-plane}}
In this confinement model, electrons are free to move along the $x$-axis but confined along $y$ and $z$ axes by square well confinement \cite{Campos1995} such that $V(y,z)=0$ for $0<y,z<b$ and $V(y,z)=\infty$ otherwise, with $b$ as the width of the square well. The effective potential is defined as
\begin{align}\label{V3}
    V_3(q)&= \frac{32}{b^{4}} \int_{0}^{b} d y \int_{0}^{b}  d y^{\prime} \, \sin ^{2}\left(\frac{\pi y}{b}\right) \sin ^{2}\left(\frac{\pi y^{\prime}}{b} \right) \nonumber\\
    &\times\int_{0}^{b} d z \int_{0}^{b}   d z^{\prime} \, \sin ^{2}\left(\frac{\pi z}{b}\right) \sin ^{2}\left(\frac{\pi z^{\prime}}{b}\right) K_{0}(q R),
\end{align}
where $R = \sqrt{(y-y^\prime)^2+(z-z^\prime)^2}$. As an exact analytic potential expression is challenging to derive, numerical results are obtained using Eq.\ (\ref{V3}), and detailed steps are given in Appendix \ref{AppendixC}. This confinement is crucial for understanding the properties of GaAs-GaAlAs quantum-well wires \cite{DRESSELHAUS20011}.

\subsubsection{\textbf{Infinite square-well potential in the \textit{y}-direction and a triangular potential in the \textit{z}-direction}}
Under this confinement scheme, electrons are trapped in the $y$-direction by square-well confinement with transverse wave function
$\phi(y)=\sqrt{\frac{2}{b}} \sin {\left(\frac{ \pi y}{b}\right)}$, while in the $z$-direction electrons experience a triangular potential $\kappa z$ for $z>0$, where $\kappa>0$ is a constant, together with hard-wall boundary conditions at $z=0$.  The solutions to the Schr\"{o}dinger equation in the $z$ direction can be expressed in terms of Airy functions; however, a simple variational \textit{Ansatz} for the lowest subband is $\zeta(z)=\frac{1}{\sqrt{2}}\left(\frac{3}{{z_0}}\right)^{3/2} \, z  \exp{\left(-\frac{3\, z}{2\,z_0}\right)} \Theta(z)$, with average wire width $z_0$ along the $z$-direction. $\Theta(z)$ is the Heaviside function. Fang and Howard \cite{fangandhoward1966negative,ando1982electronic} initially proposed this form of $\zeta(z)$ to describe the width of a two-dimensional electron gas within a semiconductor quantum well. The same wave function has been used by Cole \cite{cole1974electronic} for the lowest subband of electrons on liquid helium, provided the barrier for entering the helium is infinite and the solid-vapor interface is sharp. 
 Hence, the total wave function is $\psi(x, y, z)=\frac{e^{i k x}}{\sqrt{L}} {\left(\frac{2}{b}\right)}^{1 / 2} \sin {\left(\frac{\pi y}{b}\right)}\frac{1}{\sqrt{2}} {\left(\frac{3}{{z_0}}\right)}^{3/2} \, z  \exp{\left(-\frac{3\, z}{2\,z_0}\right)}\Theta(z)$ and it is the appropriate confinement for understanding 1D semiconductor structures \cite{Das1985}. 

The effective interaction potential can be represented as
\begin{align}\label{V4}
     V_4(q)&=  \frac{1458}{b^{2}\,{z_0}^6} \int_{0}^{b} d y \int_{0}^{b} d y^{\prime} \, \sin^2\left(\frac{\pi y}{b}\right) \sin ^{2}\left(\frac{\pi y^{\prime}}{b}\right) \nonumber \\
    &\times\int_{0}^{\infty} d z \int_{0}^{\infty} d z^{\prime} \, {z}^{2} \,z^{\prime 2} e^{- \frac{3(z + z^{\prime})}{z_0}}\,K_{0}(q R).
\end{align}

Similar to the previous confinement model, the potential cannot be expressed analytically. Equation (\ref{V4}) is used to obtain the results numerically. The details are given in Appendix \ref{AppendixD}.

\subsubsection{\textbf{Harmonic confinement in the \textit{y}-direction and a \texorpdfstring{$\bm{\sqrt{\delta(z)}}$}{sqrt{delta(z)}} wave function in the \textit{z}-direction}}
In this confinement model \cite{Hu1990}, we consider a two-dimensional electron gas in the $xy$-plane with zero thickness, in which electrons move freely in the $x$-direction with harmonic confinement in the $y$-direction. The motion of electrons in the $z$-direction is neglected. This forms a quasi-one-dimensional system as electrons occupy the lowest subband of harmonic confinement. $V_5(q)$ can be obtained analytically as
\begin{equation}\label{V5}
    V_5(q)  =\exp \left(\frac{q^{2} b^{2}}{4}\right) K_{0}\left(\frac{q^{2} b^{2}}{4}\right).
\end{equation}
The derivation of the analytic expression is presented in Appendix \ref{AppendixE}.

\subsubsection{\textbf{Infinite square-well confinement in the \textit{y}-direction and a \texorpdfstring{$\bm{\sqrt{\delta(z)}}$}{sqrt{delta(z)}} wave function in the \textit{z}-direction}}

In this case \cite{Li1989}, electrons move freely in the $x$-direction with square-well confinement in the $y$-direction, and the motion of electrons in the $z$-direction is neglected. $V_6(q)$ can be represented as
\begin{align}\label{V6}
     V_6(q)&= \frac{8}{b^2}   \int_{0}^{b} d y \int_{0}^{b} d y^{\prime} \, \sin^2\left(\frac{\pi y}{b}\right) \sin ^{2}\left(\frac{\pi y^{\prime}}{b}\right) \nonumber \\ 
& \times  K_{0}\left(q\left|y-y^{\prime}\right|\right).
\end{align}
The potential cannot be expressed analytically. Hence, Eq.\ (\ref{V6}) is calculated numerically, and the details are given in Appendix \ref{AppendixF}.

 Figure \ref{V_vs_q} depicts a significant dependence on confinement models in reciprocal space. Throughout, for the quantum wire to be 1D, the transverse dimension (wire width, $b$) must be significantly smaller than Wigner-Seitz radius $r_\text{s}$. The nonphysical situation ($b>r_\text{s}$) is where the difference between the confinement models [$V_1(q)$ to $V_6(q)$] has the greatest effect; however, the effects of the choice of confinement model on quasi-1D HEG properties can also be seen where $r_\text{s}>b$. A comparative study of analytical interelectronic interaction potentials [$V_1(q)$, $V_2(q)$, and $V_5(q)$] reveals that they exhibit similar behavior in the small-$q$ limit ($q \to 0$). Conversely, in the large-$q$ limit ($q \to \infty$), they converge to zero with distinct functional dependencies as shown in Eqs.\ (\ref{V1atlargeq}), (\ref{V2atlargeq}), and (\ref{V5atlargeq}), respectively.

\begin{figure}[htbp]
    \centering
    \includegraphics[clip,width=.48\textwidth]{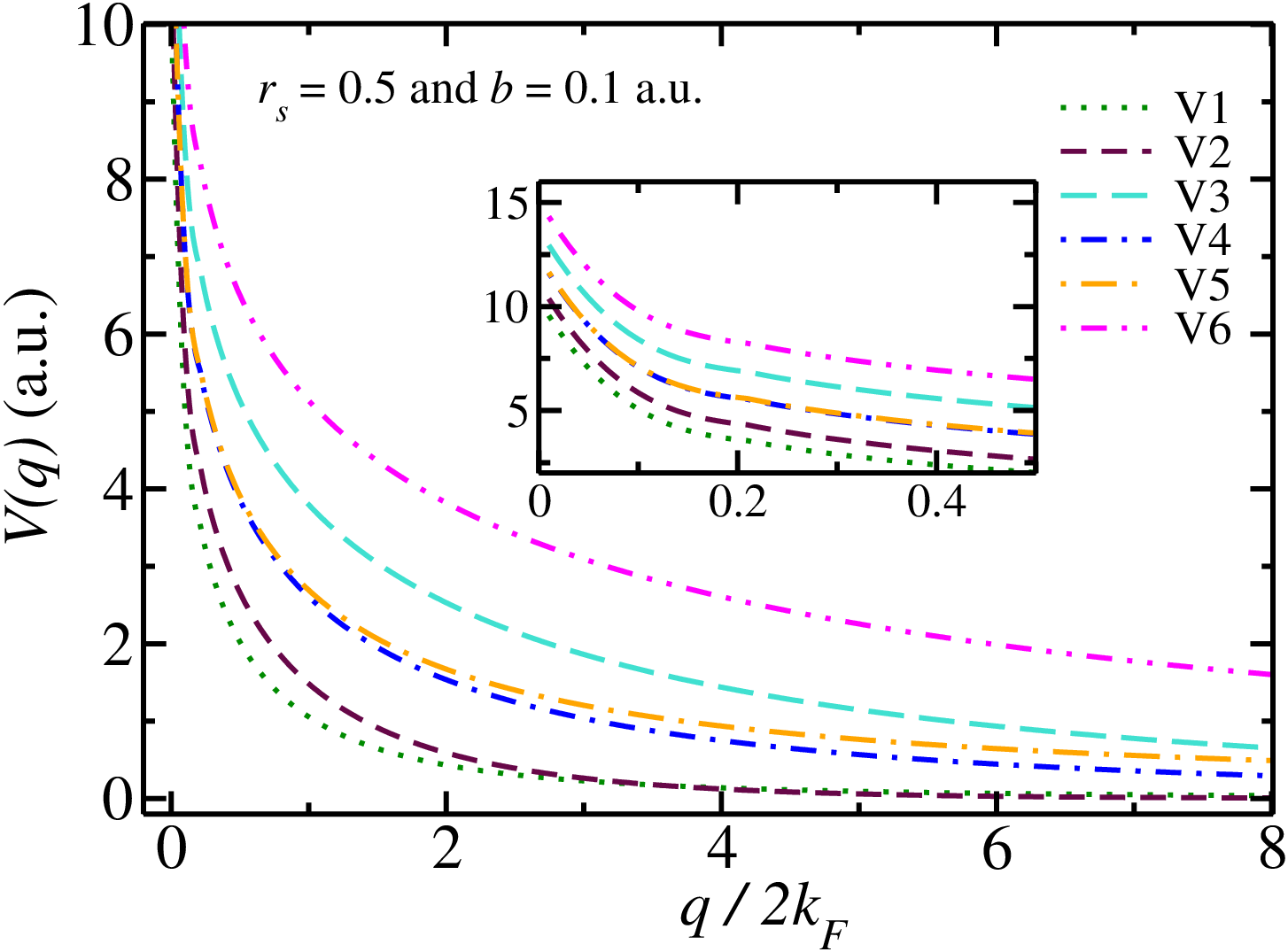}
    \caption{Effective electron-electron interaction potential $V(q)$ plotted in reciprocal space against a nondimensionalized quantity $q/2k_\text{F}$, for various confinement models. The different curves represent $V_1(q)$ to $V_6(q)$ as in Eqs.\ (\ref{V1}), (\ref{V2}), (\ref{V3}), (\ref{V4}), (\ref{V5}), and (\ref{V6}) discussed in Sec.\ \ref{models}, respectively. As shown in the inset, the potential $V(q)$
 is plotted over a small interval of $q$, demonstrating that $V_4(q)$ and $V_5(q)$ yield almost identical results in this regime. This illustrates the variation in interaction strength and range due to the geometric confinement.}
    \label{V_vs_q}
\end{figure}

% ----------------------------------------------------------------
\begin{figure*}[htbp]

\centering
\begin{minipage}[t]{0.48\textwidth}\includegraphics[clip,width=0.95\textwidth]{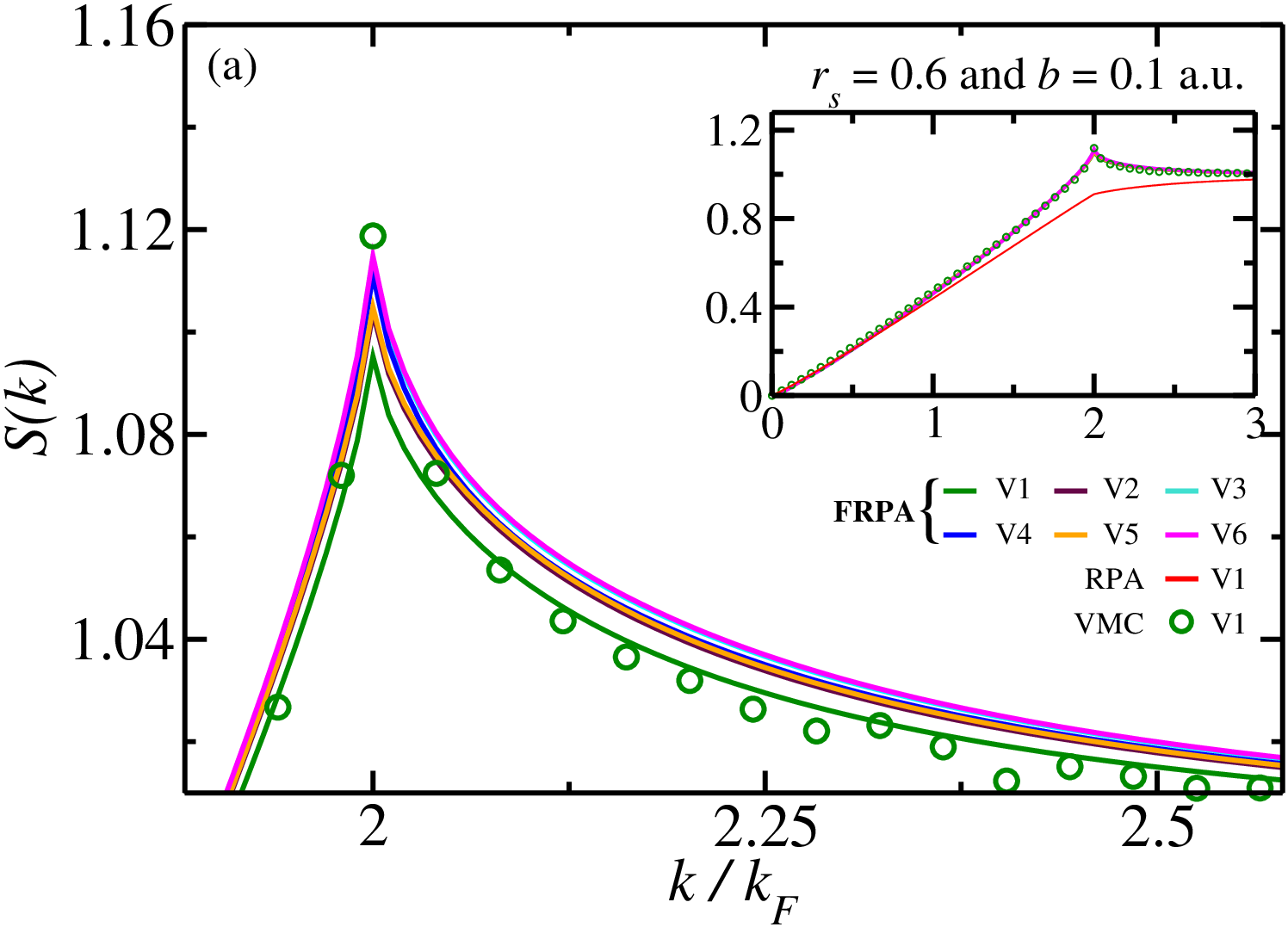}
\end{minipage}
\hfill
\begin{minipage}[t]{0.48\textwidth}
    \includegraphics[clip,width=0.95\textwidth]{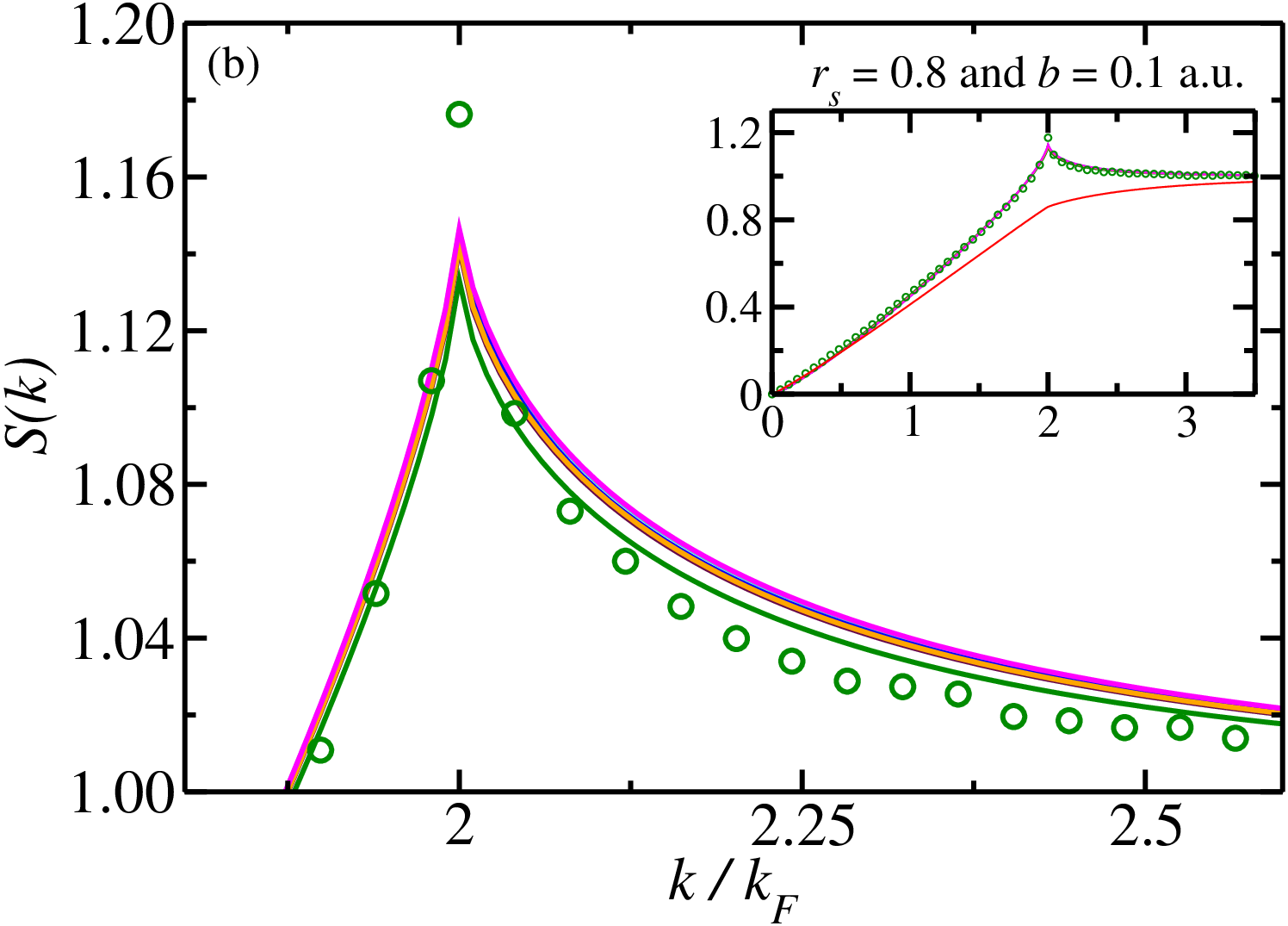}
\end{minipage}
\\[1em]
\begin{minipage}[t]{0.48\textwidth}\includegraphics[clip,width=0.95\textwidth]{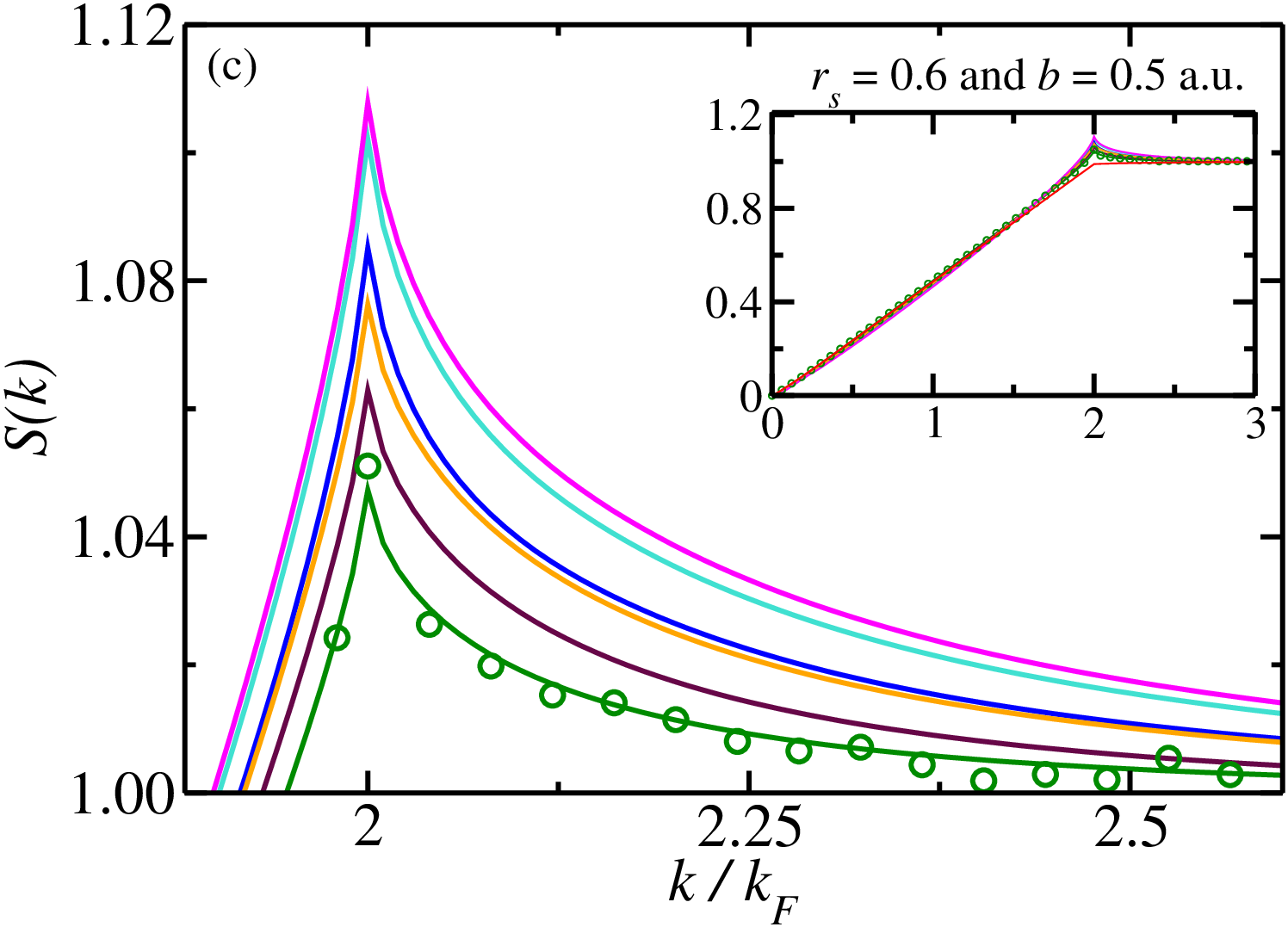}
\end{minipage}
\hfill
\begin{minipage}[t]{0.48\textwidth}
\includegraphics[clip,width=0.95\textwidth]{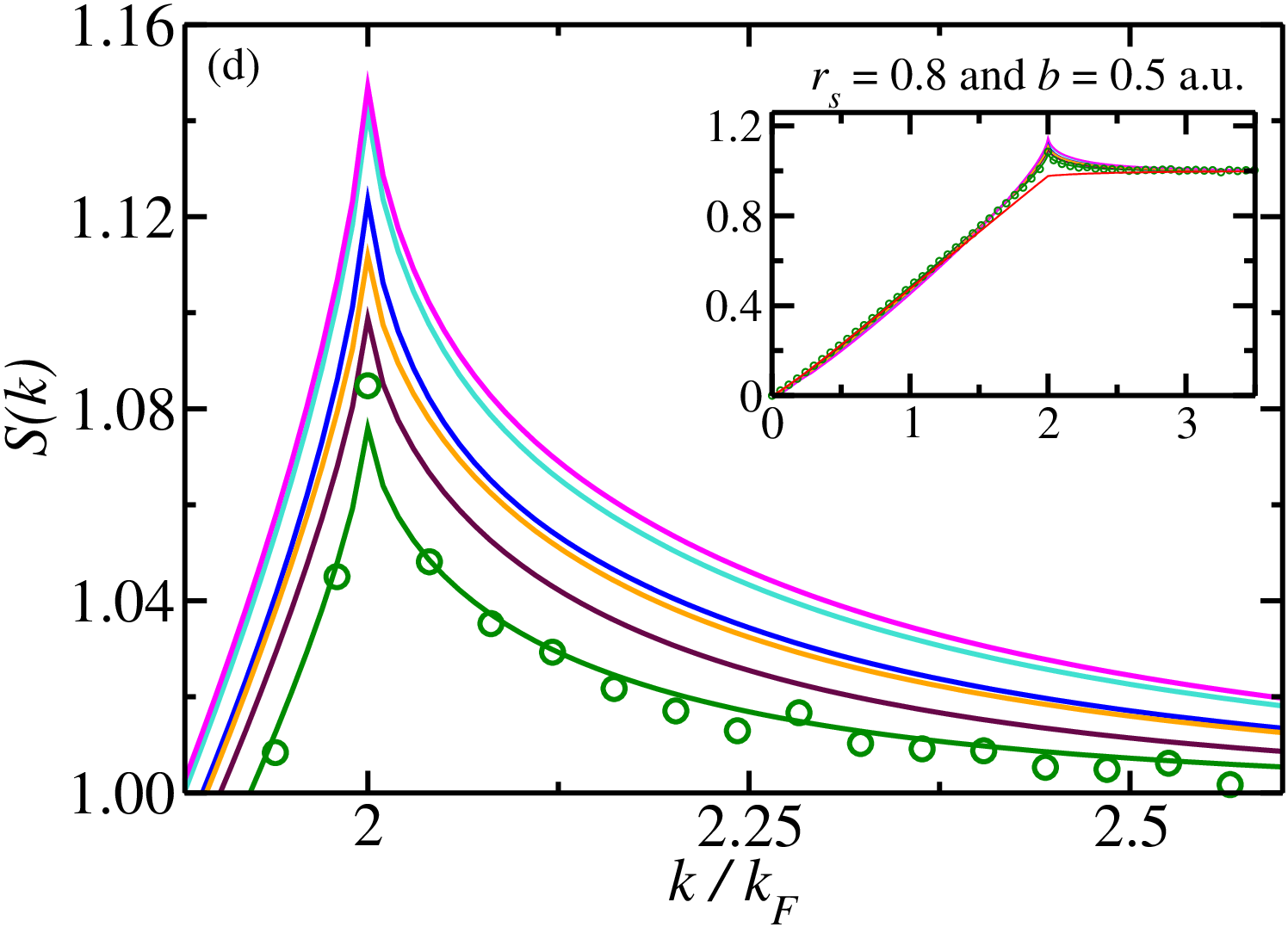}
\end{minipage}
\\[1em]
\begin{minipage}[t]{0.48\textwidth}
\includegraphics[clip,width=0.95\textwidth]{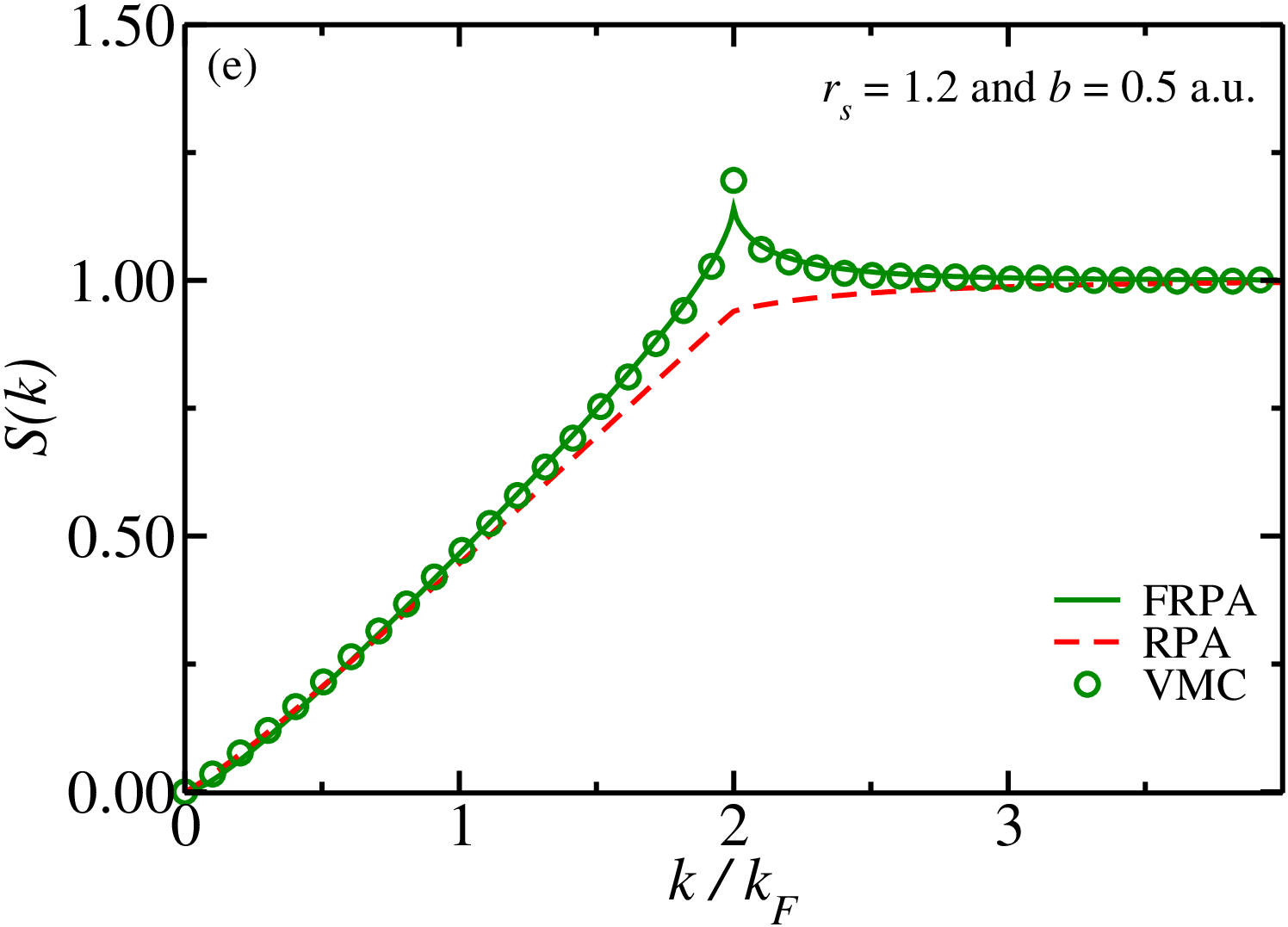}
\end{minipage}
\hfill
\begin{minipage}[t]{0.48\textwidth}
\includegraphics[clip,width=0.95\textwidth]{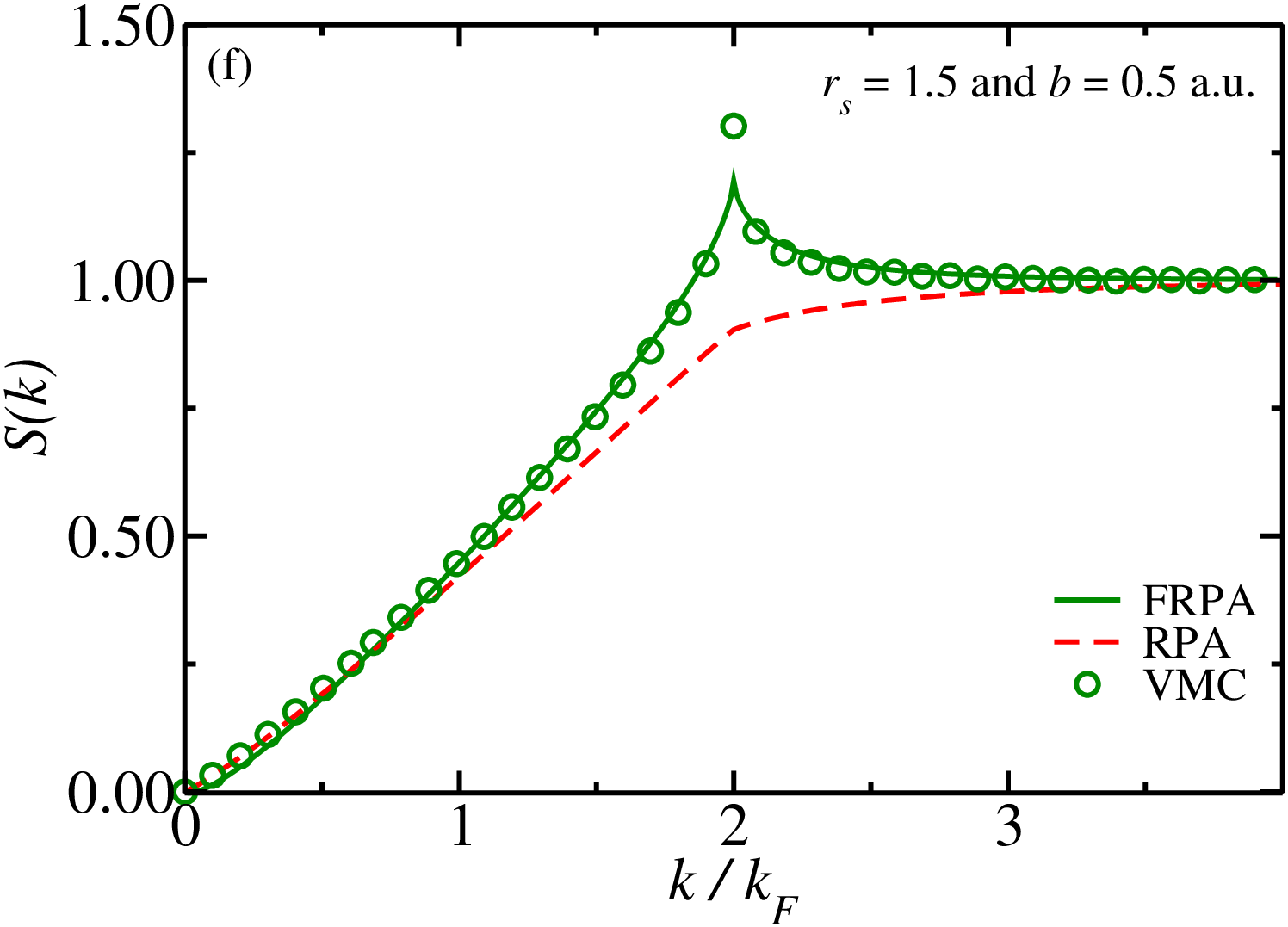}
\end{minipage}
\caption{SSF $S(k)$ plotted against $k/k_\text{F}$ for $r_\text{s} = 0.6$, 0.8, 1.2, and 1.5 with $b=0.1$ and 0.5 a.u. Using the FRPA, the SSFs for different confinement models $V_1(q)$ to $V_6(q)$ have been compared with the available VMC simulations of harmonic wires for $N=99$ electrons. In Figs.\ \ref{SSFtheoryrs_b}(a), \ref{SSFtheoryrs_b}(b), \ref{SSFtheoryrs_b}(c), and \ref{SSFtheoryrs_b}(d), the main plot shows the behavior at the $2k_\text{F}$ peak to show the variation across different confinement schemes, whereas the inset shows a zoomed-out view. In Figs.\ \ref{SSFtheoryrs_b}(e) and \ref{SSFtheoryrs_b}(f), the FRPA SSF is compared with VMC simulations and RPA SSFs.}
\label{SSFtheoryrs_b}
\end{figure*}

\begin{figure}[htbp]
\centering
\begin{minipage}[t]{0.48\textwidth}
\includegraphics[clip,width=0.95\textwidth]{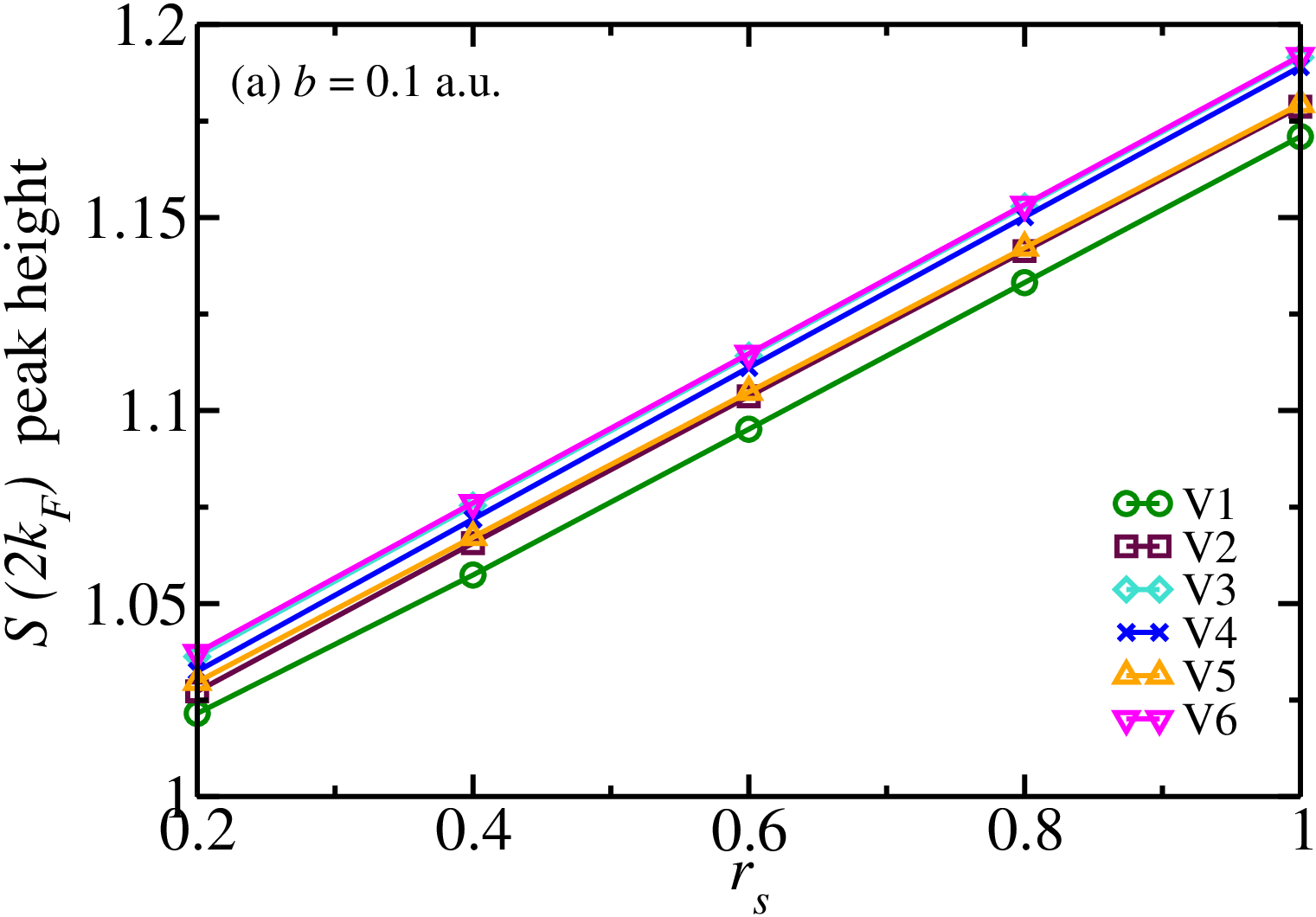}
\end{minipage}
\\[1em]
\begin{minipage}[t]{0.48\textwidth}
    \includegraphics[clip,width=0.95\textwidth]{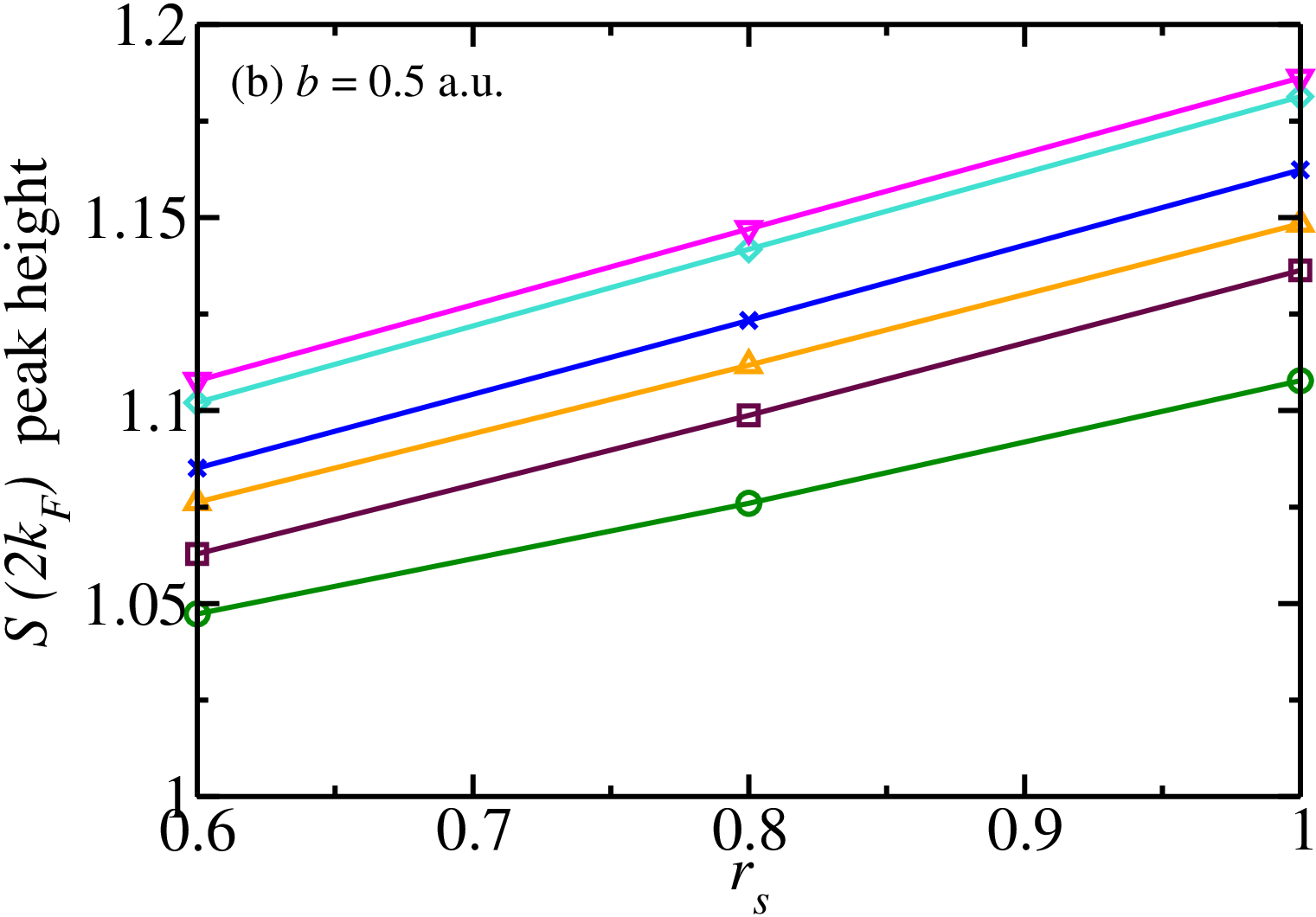}
\end{minipage}

\caption{$2k_\text{F}$ peak height of the SSF as a function of density parameter $r_\text{s}$ for different confinement models [$V_1(q)$--$V_6(q)$] and (a) $b=0.1$ and (b) $b=0.5$ a.u. The data points representing the peak heights are joined by lines, revealing a linear relationship with $r_\text{s}$ at fixed wire width ($b<r_\text{s}$).}
\label{lssf_peaks_vs_rs_bfixed}
\end{figure}
% ----------------------------------------------------------------
\begin{figure}[t]
\centering
\begin{minipage}[t]{0.48\textwidth}
\includegraphics[clip,width=0.95\textwidth]{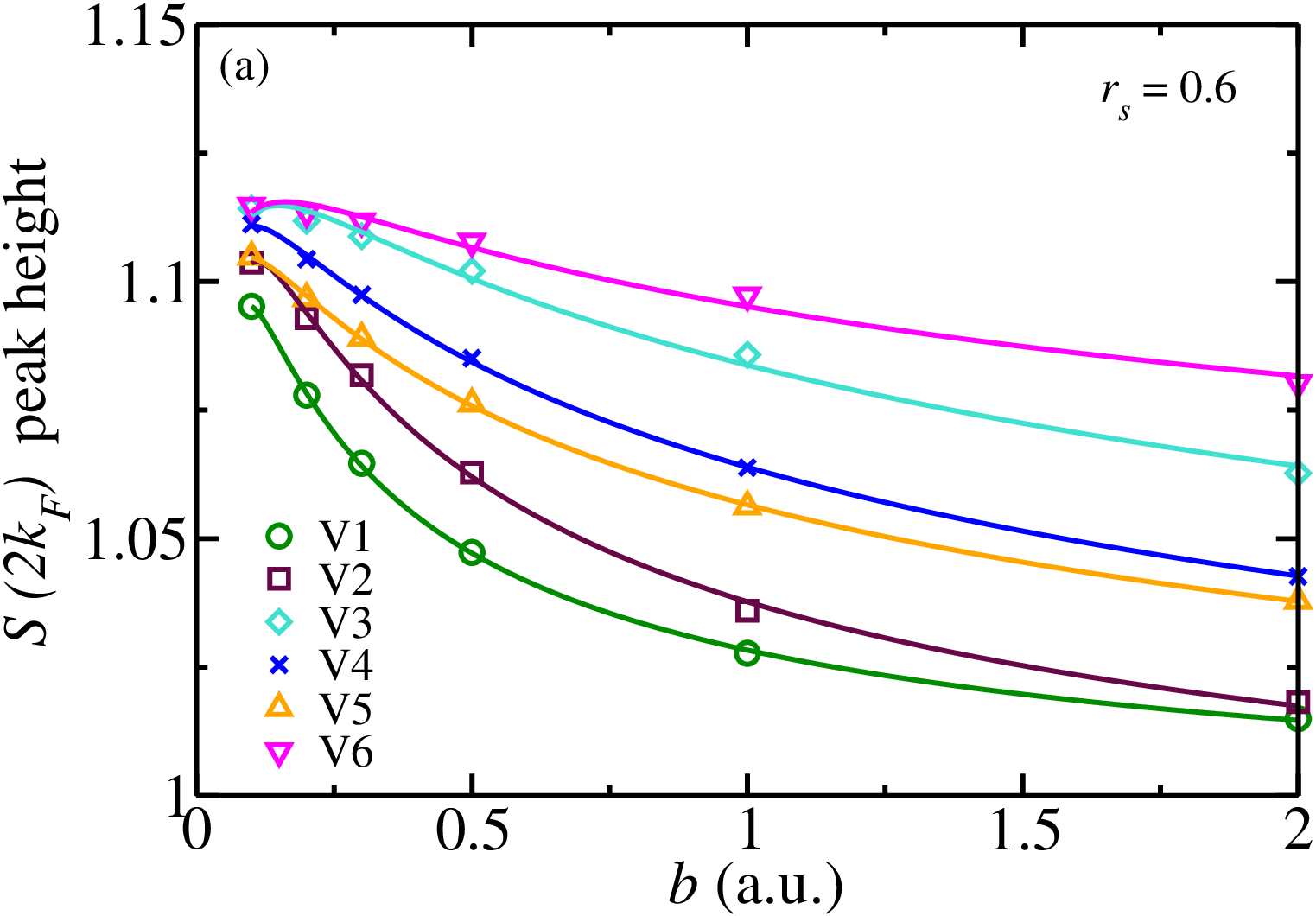}
\end{minipage}
\\[1em]
\begin{minipage}[t]{0.48\textwidth}
\includegraphics[clip,width=0.95\textwidth]{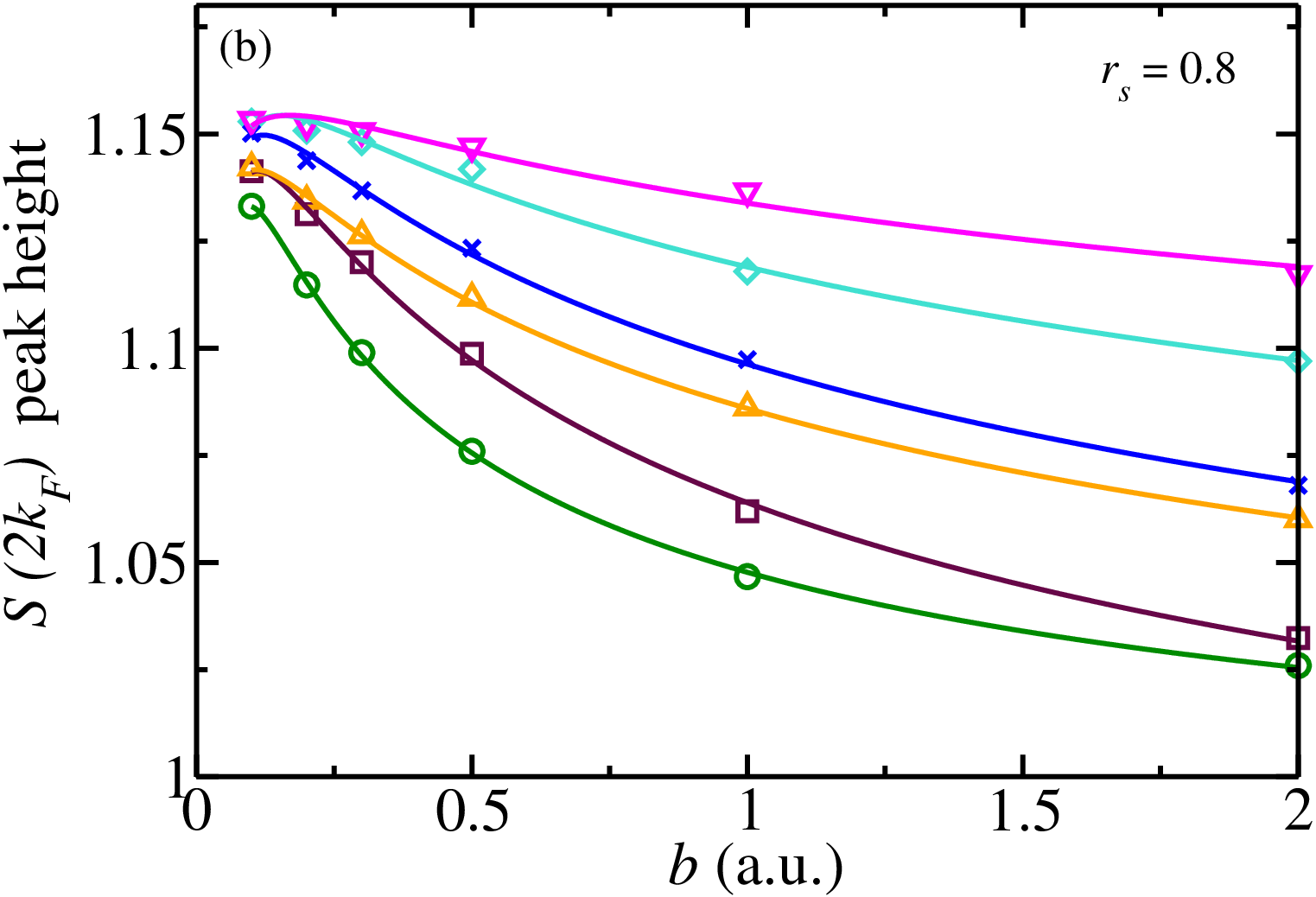}
\end{minipage}

\caption{$2k_\text{F}$ peak height of the SSF, fitted by Eq.\ (\ref{Eq:SSF2kf_fit}), for different confinement models [$V_1(q)$--$V_6(q)$]. The different symbols represent our FRPA data, and the solid lines show the corresponding fitted function for (a) $r_\text{s} = 0.6$ and (b) $r_\text{s} = 0.8$.} 
\label{lssf_peaks_vs_b_rsfixed}
\end{figure}
% ----------------------------------------------------------------

\begin{figure*} 
\centering
\begin{minipage}[t]{0.48\textwidth}
    \includegraphics[clip,width=0.95\textwidth]{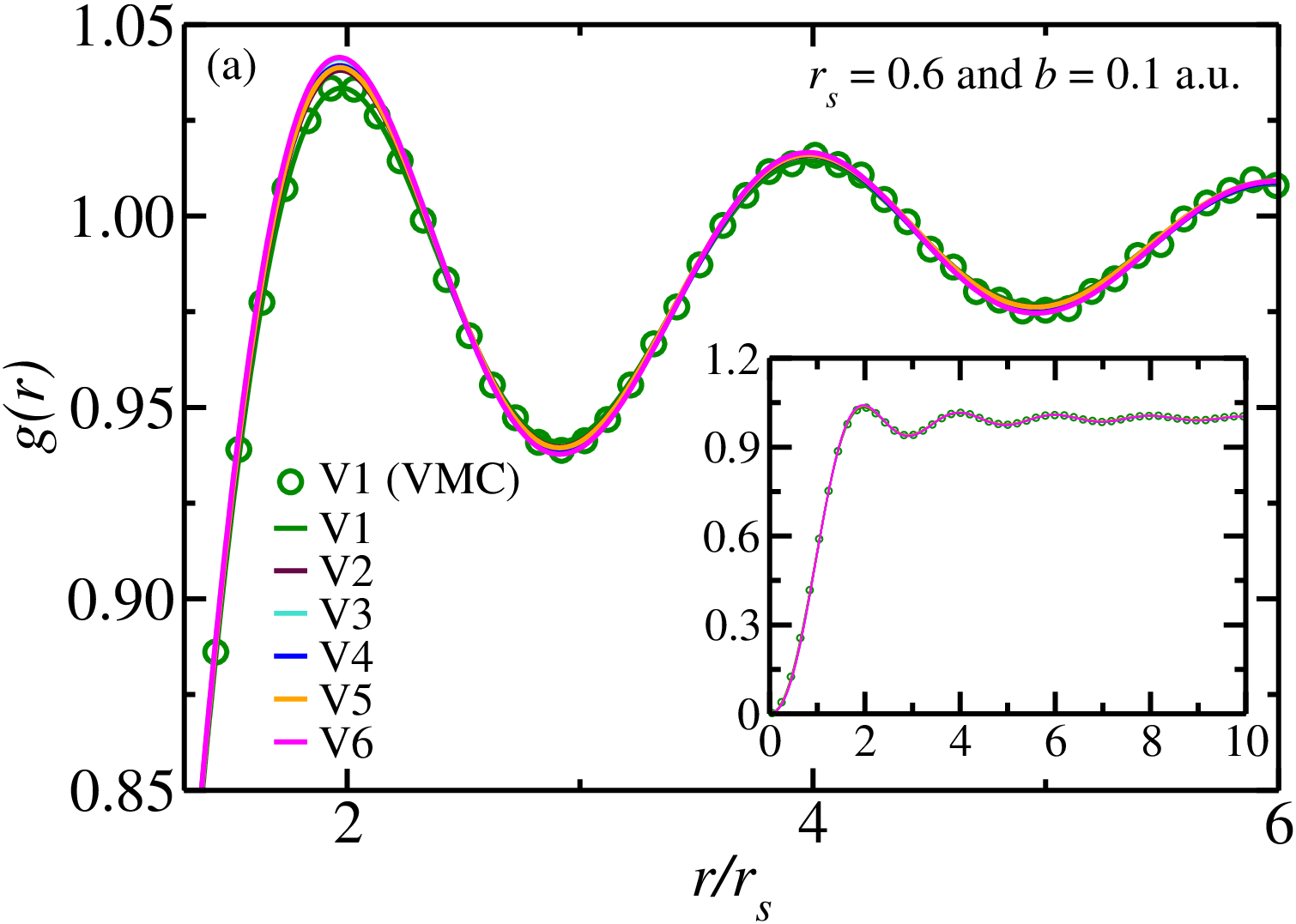}
\end{minipage}
\hfill
\begin{minipage}[t]{0.48\textwidth}
    \includegraphics[clip,width=0.95\textwidth]{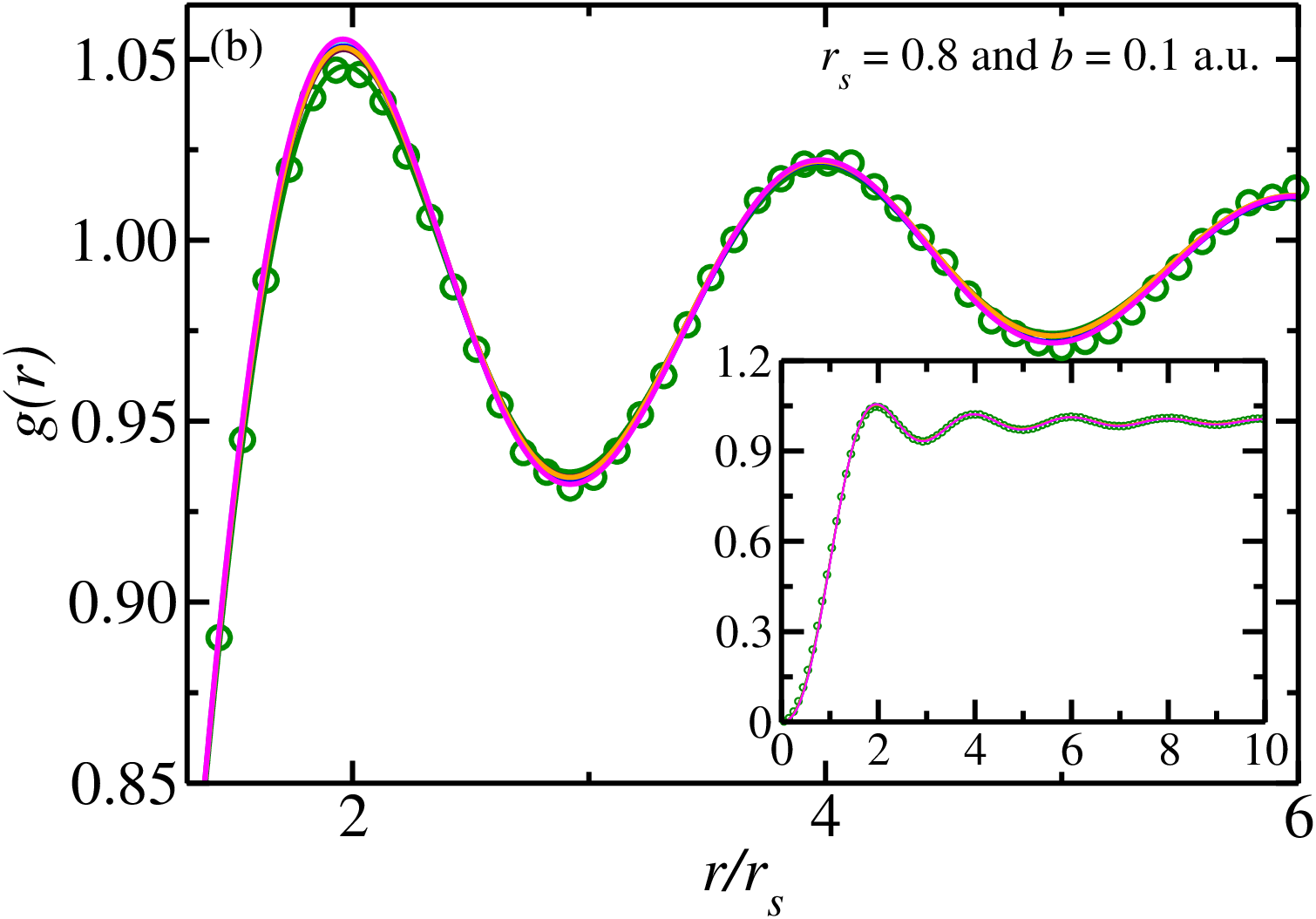}
\end{minipage}
\\[1em]
\begin{minipage}[t]{0.48\textwidth}
\includegraphics[clip,width=0.95\textwidth]{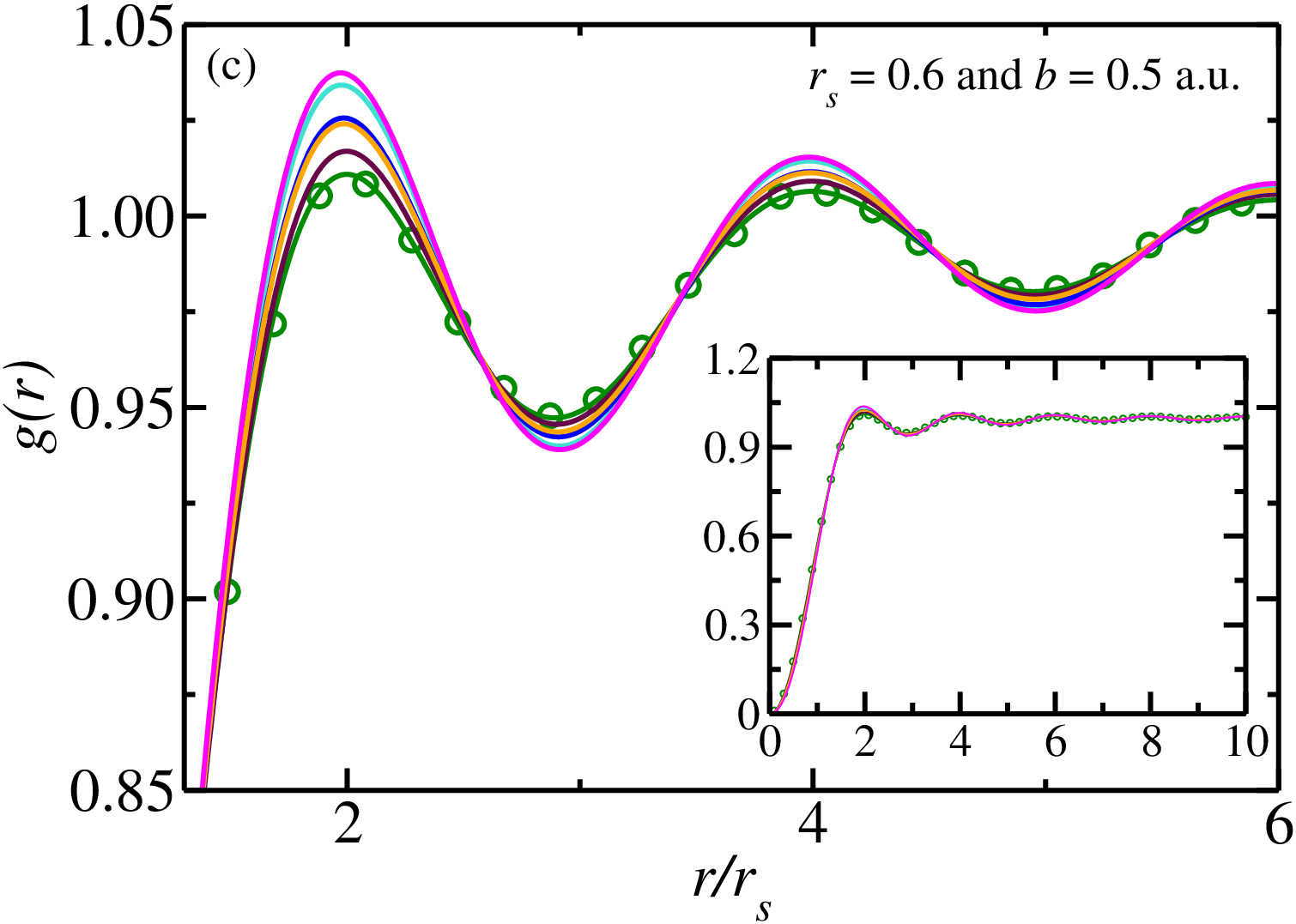}
\end{minipage}
\hfill
\begin{minipage}[t]{0.48\textwidth}
\includegraphics[clip,width=0.95\textwidth]{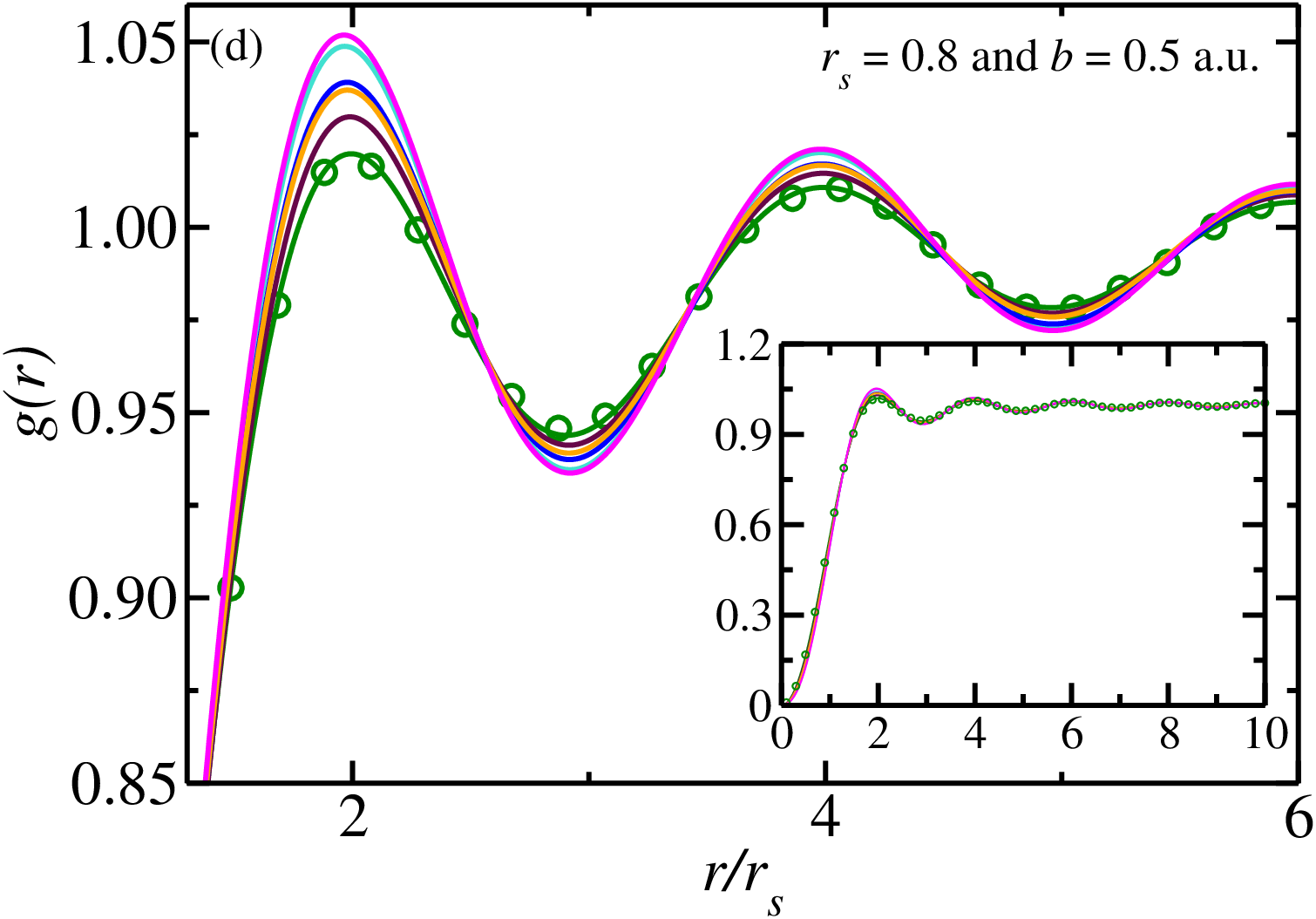}
\end{minipage}
\\[1em]
\begin{minipage}[t]{0.48\textwidth}
\includegraphics[clip,width=0.95\textwidth]{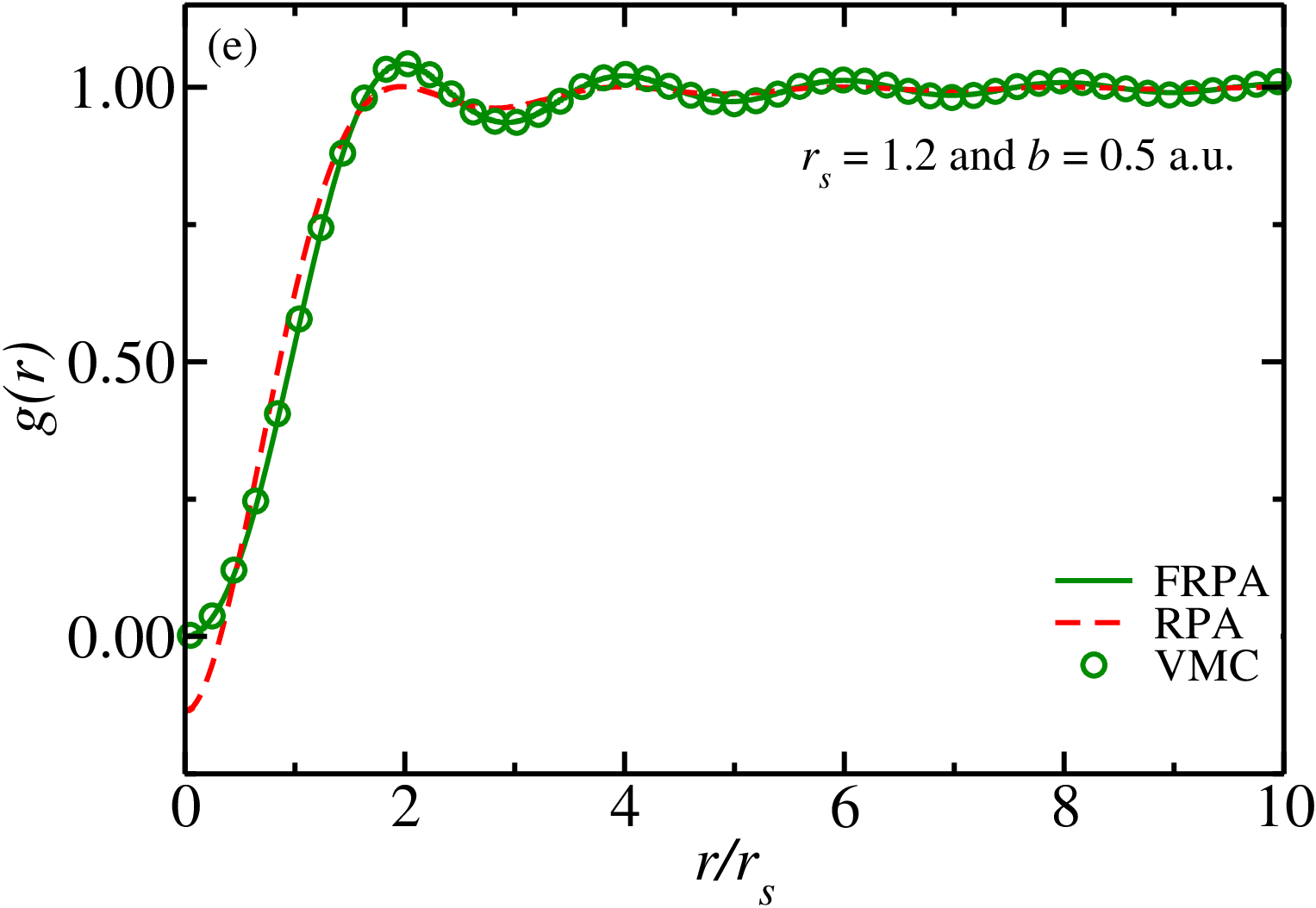}
\end{minipage}
\hfill
\begin{minipage}[t]{0.48\textwidth}
\includegraphics[clip,width=0.95\textwidth]{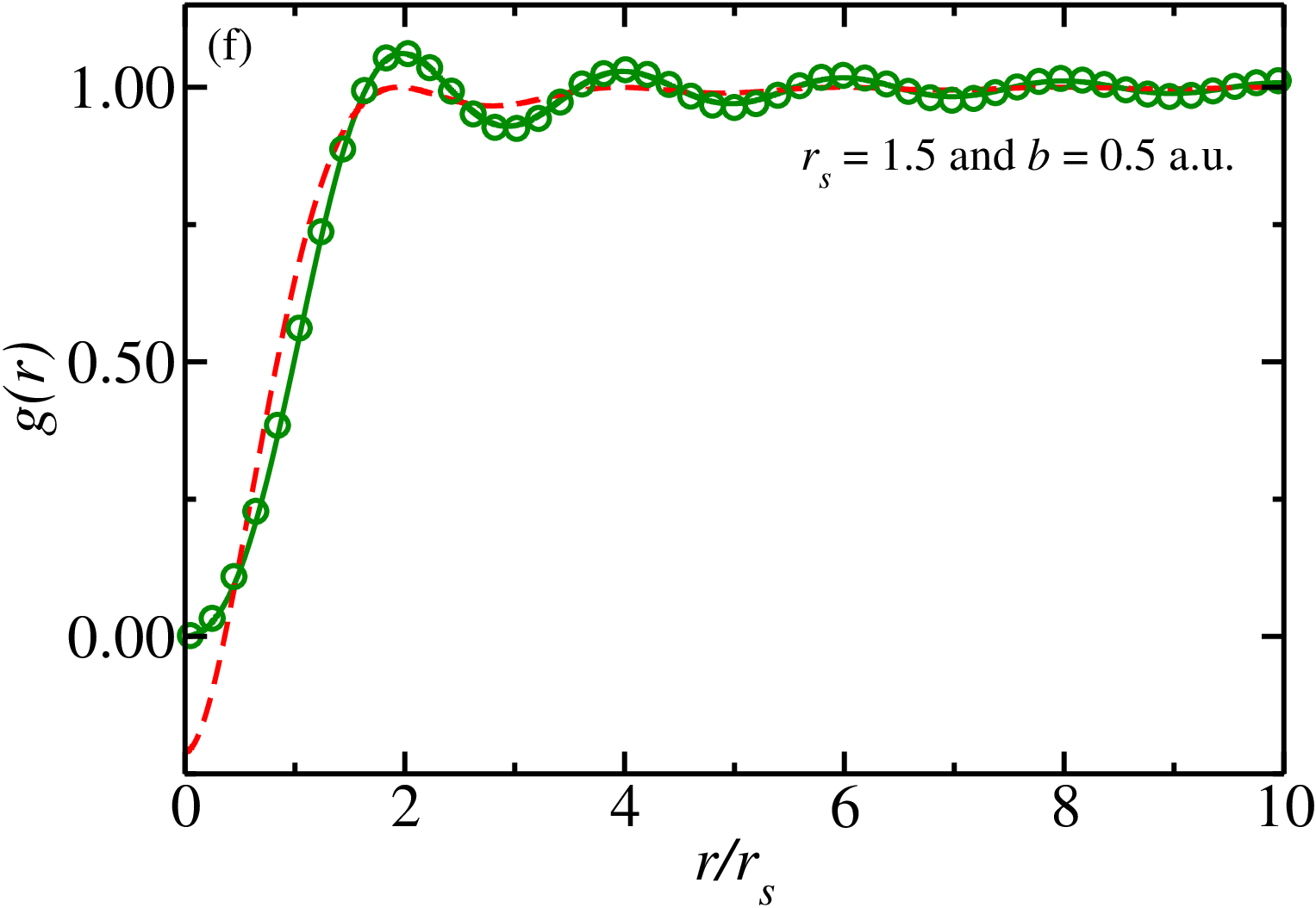}
\end{minipage}
\caption{PCF plotted against $r/r_\text{s}$ for $r_\text{s}=0.6$, 0.8, 1.2, and 1.5 with $b=0.1$ and 0.5 a.u.  The PCFs for different confinement models $V_1(q)$ to $V_6(q)$ have been compared with the available VMC simulations of harmonic wires with $N = 99$ electrons. In Figs.\ \ref{PCFtheory_rs_b}(a), \ref{PCFtheory_rs_b}(b), \ref{PCFtheory_rs_b}(c), and \ref{PCFtheory_rs_b}(d), the main plot shows a zoomed-in view of the amplitude of oscillations around $r = 2r_\text{s}$, which varies depending on the coupling parameters ($b$, $r_\text{s}$) and the confinement model, whereas the inset shows a zoomed-out view. In Figs.\ \ref{PCFtheory_rs_b}(e) and \ref{PCFtheory_rs_b}(f), the FRPA PCF is compared with VMC simulations and the RPA\@.}
\label{PCFtheory_rs_b}
\end{figure*}

\subsection{Density response function}

In this section, we present an analysis of the static properties of the 1D HEG using the dynamical density response function and the fluctuation-dissipation theorem \cite{Giuliani2008}. The density response function quantifies the change in electron density due to a change in the external potential. It is obtained in FRPA and given as \cite{bala2012exchange}
\begin{equation}
    \label{drf_exact}
    \hspace{-0.15cm}\chi(q, \omega)=  \frac{\chi_{0}(q, \omega)+\lambda \{ \chi_{1}^\text{se}(q, \omega) +\chi_{1}^\text{ex}(q, \omega) \}}{1-\lambda V(q)\left[\chi_{0}(q, \omega)+\lambda \{ \chi_{1}^\text{se}(q, \omega) +\chi_{1}^\text{ex}(q, \omega) \}\right]},
\end{equation}
where $V(q)$ is the Fourier transform of the effective potential and $\lambda$ denotes the order of the potential, while $\chi_{1}^\text{se}(q, \omega)$ and $\chi_{1}^\text{ex}(q, \omega)$ represent the first-order self-energy and exchange correction to the polarizability, respectively.
In the weak coupling ($\approx$ high density) approximation, the FRPA density response function, Eq.\ (\ref{drf_exact}), can be written as
\begin{eqnarray}
 \label{resHDE}
 \chi(q,\omega)&\approx&\chi_{0}(q,\omega)+\lambda\; V(q)\chi_{0}^2(q,\omega)\nonumber\\
 & &+\lambda\; \chi_{1}^\text{se}(q,\omega)+\lambda \;\chi_{1}^\text{ex}(q,\omega).
\end{eqnarray}
This approximation enables us to do most of the calculations analytically \cite{ashokan2020exact}.
The noninteracting polarizability
 describes the response of a system of independent particles to an external perturbation and is simplified as
\begin{eqnarray}\label{chi0}
 \chi_{0}(q, \omega)=\frac{g_{s} m}{2 \pi q}
 \ln \bigg[\frac{\omega^2-(\frac{q^2}{2 m}-\frac{q
   k_\text{F}}{m})^2}{\omega^2-(\frac{q^2}{2 m}+ \frac{q k_\text{F}}{m})^2}\bigg].
\end{eqnarray}

The simplified forms of the self-energy and exchange contributions are, respectively, \cite{ashokan2020exact,girdhar2022electron}
\begin{eqnarray}
   \chi_{1}^\text{se}(q,\omega)&&=2g_\text{s}\sum_{k,p}n_k n_p [V(k-p)-V(k-p+q)]\nonumber\\
  &&\times \frac{\Omega^2_{k,q}+\omega^2}{(\Omega^2_{k,q}-\omega^2)^2}\label{chi1self}
\end{eqnarray}
and
\begin{eqnarray}
    \chi_{1}^\text{ex}(q,\omega)&&=-2g_\text{s}\sum_{k,p}\bigg\{V(k\!-\!p) [n_{k-\frac{q}{2}}n_{p-\frac{q}{2}}\!-\!n_{k-\frac{q}{2}}n_{p+\frac{q}{2}}]\nonumber\\
  &&\times\frac{(\Omega_{k-\frac{q}{2},q}\;\Omega_{p-\frac{q}{2},q}+ \omega^2)}{(\Omega^2_{k-\frac{q}{2},q}-\omega^2)(\Omega^2_{p-\frac{q}{2},q}- \omega^2)}\bigg\}
\end{eqnarray}
with 
\begin{eqnarray}
   \Omega_{k,q}=\omega_k-\omega_{k+q}\,\,\,\,\,\,\,;\,\,\,\,\,\,\,\Omega_{p,q}=\omega_p-\omega_{p+q},
\end{eqnarray}
where $g_\text{s}$  is defined as the spin degeneracy factor and $n_k$ represents the Fermi-Dirac distribution function. The above expressions are directly incorporated into our calculations.

% ------------------------------------------------------------------------------
\section{SSF}\label{SSF}
The SSF is defined as
\begin{eqnarray}
 \label{ssf}
 S(q)=-\frac{1}{n\pi} \int_{0}^{\infty} d\omega\; \chi''(q,\,
 \omega),
\end{eqnarray}
where $\chi''(q,\omega)$ corresponds to the imaginary part of the density response function, and $n=(k_\text{F}~g_\text{s})/\pi$ is the number density of electrons. 
To evaluate the integral in Eq.\ (\ref{ssf}), we employ the contour integration technique \cite{Giuliani2008} as
\begin{eqnarray}
 \label{ssfRPA}
 S(q)=-\frac{1}{n\pi} \int_{0}^{\infty} d\omega\; \chi(q,i\,
 \omega).
\end{eqnarray}
Substituting Eq.\ (\ref{resHDE}) into Eq.\ (\ref{ssfRPA}), the total SSF can be expressed as
\begin{eqnarray}
 \label{ssftotal}
 S(q)=S_0(q)+S^\text{d}_1(q)+S^\text{se}_1(q)+S^\text{ex}_1(q),
\end{eqnarray}
where 
\begin{eqnarray}
 S_0(q)&=& -\frac{1}{n\pi}\int^\infty_0  \chi_0(q,i\omega) \, d\omega \label{noninteractingSSFS0q}\\
S_1^\text{d}(q)&=& -\frac{1}{n\pi}\int^\infty_0  V(q) \chi_0^2(q,i\omega) \, d\omega\\
S_1^\text{se}(q)&=& -\frac{1}{n\pi}\int^\infty_0 \chi_1^\text{se}(q,i\omega) \, d\omega \label{S1self}
% \label{Sse_d}
\\
S_1^\text{ex}(q)&=& -\frac{1}{n\pi}\int^\infty_0 \chi_1^\text{ex}(q,i\omega) \, d\omega.
\end{eqnarray}
   
The noninteracting SSF is calculated using  Eq.\ (\ref{chi0}) and Eq.\ (\ref{noninteractingSSFS0q}),  simplifies as \cite{ashokan2018dependence}
\begin{equation}
S_0(x) = 
\begin{cases}
x, & x < 1 \\
1, & x > 1  \,\,\,\,\,\text{with}\,\,\,\,\,\, x = q/2k_\text{F}
\end{cases}.
% \label{S0}
\end{equation}
The first-order SSF can also be expressed as
\begin{equation} 
    S_1(x)=S_1^\text{d}(x)+S_1^\text{se}(x)+S_1^\text{ex}(x).
    % \label{ssf1}
\end{equation}
$S_{1}^\text{se}(x)$ does not contribute to the SSF as the integral of the self-energy term over $\omega$ vanishes as shown in Appendix \ref{AppendixG}. There are contributions to the SSF only from the exchange and direct terms. The analytical expressions of $S_1^\text{d}(x)$ and $S_1^\text{ex}(x)$ have been reported in the literatures \cite{ashokan2020exact,KM2018,girdhar2022electron}. These are presented below for ready reference. In this paper, we will numerically evaluate $S_1(x)$  for other confinement models using the following expressions.

The contribution of direct term to the SSF \cite{ashokan2018one,KM2018} is obtained for $x<1$ as
\begin{align}
  S_1^\text{d}(x)&=-\frac{g_\text{s}^2 r_\text{s}} {\pi ^2 x} \bigg[ \bigg((1-x) \ln (1-x)\nonumber\\
  &+(x+1) \ln (x+1)\bigg)V(x)\bigg]
\end{align}
and similarly for $x>1$, 
 \begin{align}
 S_1^\text{d}(x)&=-\frac{g_\text{s}^2 r_\text{s}}{\pi ^2 x}\bigg[\bigg( (x-1) \ln (x-1)-2 x \ln (x)\nonumber\\
 &+(x+1) \ln (x+1)\bigg)V(x) \bigg].
\end{align}
The contribution of the exchange term to the SSF \cite{ashokan2020exact} is obtained for $x<1$ as 
\begin{align}
 S^\text{ex}_1(x)&=\frac{g_\text{s}^2r_\text{s}}{\pi^2 x}
 \bigg[\left ((1+x)\int\limits_{1}^{1+x}-(1-x)\int\limits_{1-x}^{1}\right )
 \frac{d \bar x}{ \bar x} \, {V}(\bar x) \nonumber\\
 &+\left (\int\limits_{1-x}^{1} - \int\limits_{1}^{1+x~}\right )
 d {\bar x} V(\bar x)\bigg]
 \label{Sexa}
\end{align}
and similarly for $x>1$, 
\begin{align}
 S^\text{ex}_1(x)&=\frac{g_\text{s}^2r_\text{s}}{\pi^2 x}
 \bigg[\left ( (1+x)\int\limits_{x}^{1+x}-
 (x-1)\int\limits_{x-1}^{x}\right ) \frac{d \bar x}{ \bar x} \, {V}(\bar x)\nonumber\\
 &+\left (\int\limits_{~x-1}^{x} - \int\limits_{x}^{1+x~}\right )
 d {\bar x} \, V(\bar x)\bigg].
 \label{Sexb}
\end{align}

The exchange term can be calculated numerically by solving the integrals in Eqs.\ (\ref{Sexa}) and (\ref{Sexb}), however the integral can only be solved analytically for harmonic and cylindrical confinement potentials. It does not require discussion here as it has been reported previously in the literature \cite{girdhar2022electron,ashokan2020exact}.

In Fig.\ \ref{SSFtheoryrs_b}, the SSF is numerically evaluated using Eq.\ (\ref{ssftotal}) and plotted for $r_\text{s} =0.6$, 0.8, 1.2, and 1.5 with $b=0.1$ and $0.5$ a.u.
The SSF exhibits a prominent peak at $2k_\text{F}$; a magnified view can be seen in the main plot, whereas the zoomed-out plot is shown in the inset of the figure. 
The $2k_\text{F}$ peak heights in the SSF are plotted as a function of the density parameter ($r_\text{s}$) and wire width ($b$) in Figs.\ \ref{lssf_peaks_vs_rs_bfixed} and \ref{lssf_peaks_vs_b_rsfixed}, respectively. 
% ----------------------------------------------------------------
\section{PCF}\label{PCF}
The PCF $g(r)$ is obtained by performing an inverse Fourier transform on the SSF $S(q)$ as
\begin{eqnarray}
 \label{pcf_formula}
 g(r)=1-\frac{1}{2 n\pi}\int^{\infty}_{-\infty} dq \, e^{iqr}[1-S(q)].
\end{eqnarray}

It quantifies the probability of finding a particle at a certain distance from another in real space.
Using Eq.\ (\ref{pcf_formula}), the PCFs for various confinement schemes are evaluated and subsequently plotted in Fig.\ \ref{PCFtheory_rs_b} for various density parameters with wire widths $b=0.1$ and $0.5$ a.u.

\begin{figure}[b]
\centering
\begin{minipage}[t]{0.48\textwidth}
\includegraphics[clip,width=0.95\textwidth]{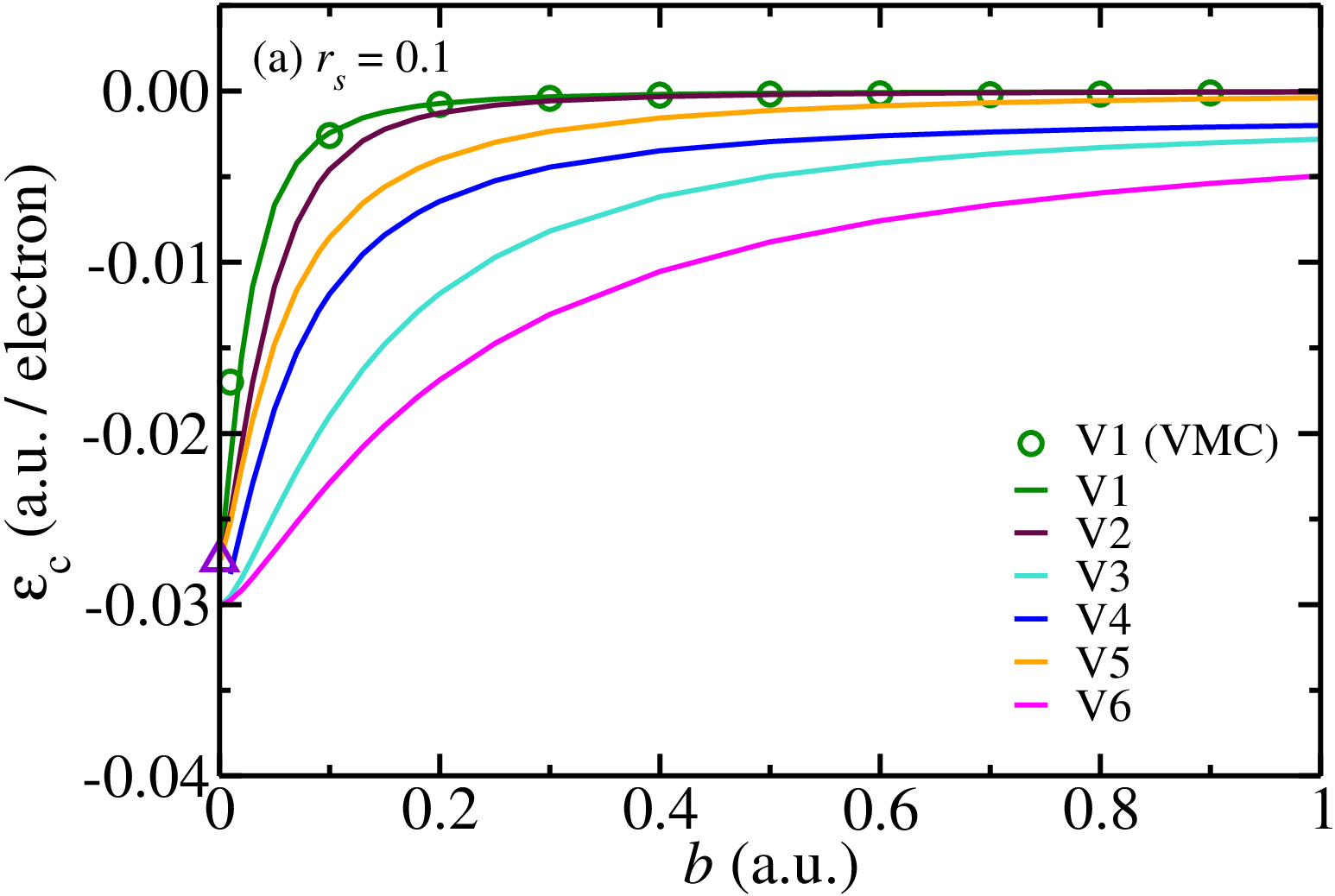}
\end{minipage}
\\[1em]
\begin{minipage}[t]{0.48\textwidth}
\includegraphics[clip,width=0.95\textwidth]{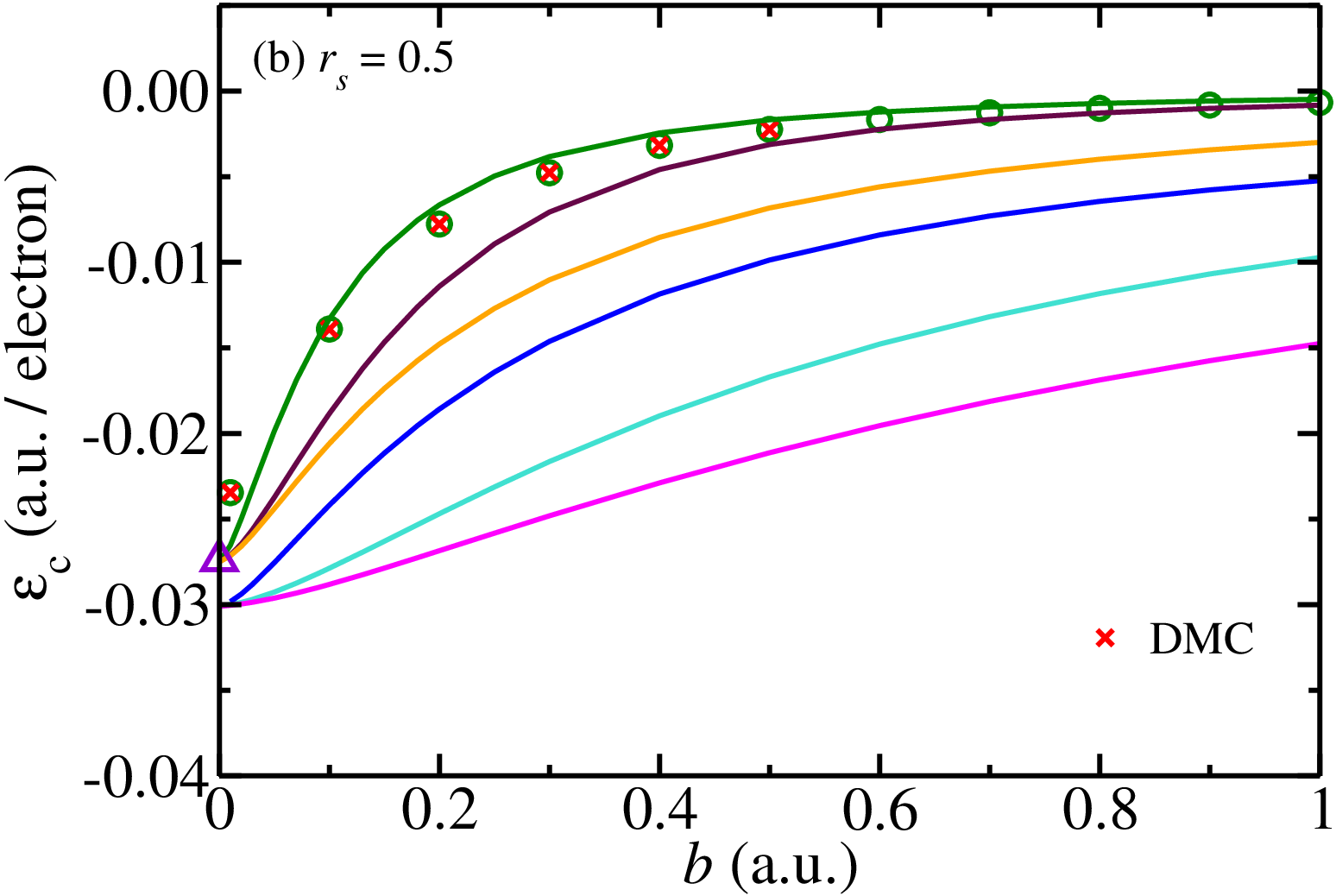}
\end{minipage}

\caption{Correlation energy per electron ($\epsilon_\text{c}$) plotted as a function of wire width $b$ for $r_\text{s}=0.1$ and $0.5$. The different confinement models $V_1(q)$ to $V_6(q)$ have been compared with the VMC- and DMC-simulated correlation energy of harmonic wire for $r_\text{s}=0.1$ and $0.5$ \cite{ashokan2018one,girdhar2022electron}. The triangle symbol $\textcolor{violet}{\boldsymbol{\Delta}}$ indicates the correlation energy $\epsilon_\text{c} =-\pi^2/360$ at the high-density limit for an ultrathin wire. The correlation energy for the confinement models $V_1(q)$, $V_2(q)$, and $V_5(q)$ approaches $\epsilon_\text{c}(r_\text{s})= - \pi^2/360$, whereas other confinement potentials $V_3(q)$, $V_4(q)$, and $V_6(q)$ do not approach the same high-density limit.
\cite{ashokan2020exact}.}
\label{lce_vs_b_rsfixed}
\end{figure}
% ----------------------------------------------------------------
\section{Ground state energy}\label{Groundstateenergy}
The ground-state energy can be determined from the density-density response function and fluctuation-dissipation theorem as \cite{ashokan2020exact},
\begin{eqnarray}
\label{gsE}
E_\text{g}&=&E_0+\frac{n}{2}\sum_{q\neq 0}V(q)\nonumber\\
& \times&\bigg( -\frac{1}{n\pi}\int^1_0 d\lambda \int^{\infty}_0 \chi(q,i \omega;\lambda)\; d\omega-1\bigg).
\end{eqnarray}
Substituting Eq.\ (\ref{resHDE}) into Eq.\ (\ref{gsE}), the ground-state energy of the HEG can be decomposed into  three components: the noninteracting kinetic energy $E_0$, the exchange energy $E_\text{x}$, and the correlation energy $E_\text{c}$ as
\begin{eqnarray}
 \label{gse_all}
 E_\text{g}=E_0+E_\text{x}+E_\text{c}.
\end{eqnarray}

For the fully spin-polarized ($k_\text{F}= \pi/2r_\text{s}$) HEG, integrating the kinetic energy of each electron over all occupied states yields the total kinetic energy, which is given by $E_0={\pi ^2}/{24 r_\text{s}^2}$. The exchange energy contribution can be expressed as
\begin{eqnarray}
E_\text{x}&=&\frac{n}{2}\sum_{q\neq 0}V(q)\bigg( -\frac{1}{n\pi}\int^1_0 d\lambda \int^\infty_0 \chi_0(q,i\omega) \, d\omega-1\bigg)\nonumber\\
&=&\frac{n}{2}\sum_{q\neq 0}V(q) [S_0(q)-1]
\end{eqnarray}
and the correlation energy is given by
\begin{eqnarray}
E_\text{c}&=&\frac{n}{2}\sum_{q\neq 0}V(q)\bigg [ -\frac{1}{n\pi}\int^1_0 d\lambda \int^\infty_0 \bigg ( \lambda\;V(q) \chi_0^2(q,i\omega)\nonumber\\
& +&\lambda\; \chi_1^\text{se}(q,i\omega)+\lambda\; \chi_1^\text{ex}(q,i\omega)\bigg ) \, d\omega\bigg ]\\
&=&\frac{n}{4}\sum_{q\neq 0}V(q) [S^\text{d}_1(q)+S_1^\text{se}(q)+S_1^\text{ex}(q)].
\end{eqnarray}

The correlation energy per particle can be expressed as
\begin{equation}
    \epsilon_\text{c} = \frac{1}{4\pi}\int_{0}^{\infty} V(q) [S^\text{d}_1(q)+S_1^\text{se}(q)+S_1^\text{ex}(q)] \, dq.
    \label{eq:corr_e_analytic}
\end{equation}

\begin{figure}[htbp]
\centering
\begin{minipage}[t]{0.48\textwidth}
    \includegraphics[clip,width=0.95\textwidth]{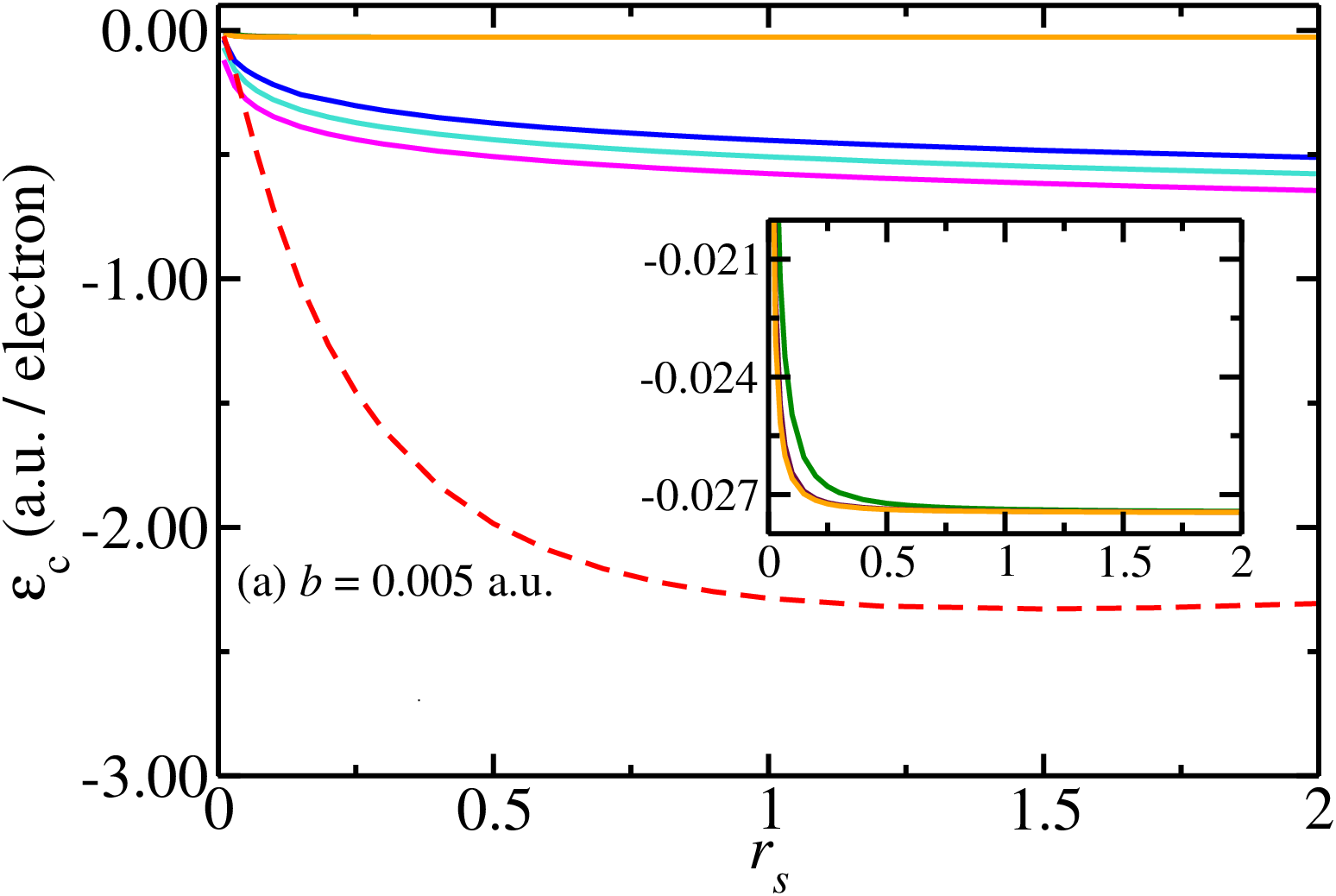}
\end{minipage}
\\[1em]
\begin{minipage}[t]{0.48\textwidth}
    \includegraphics[clip,width=0.95\textwidth]{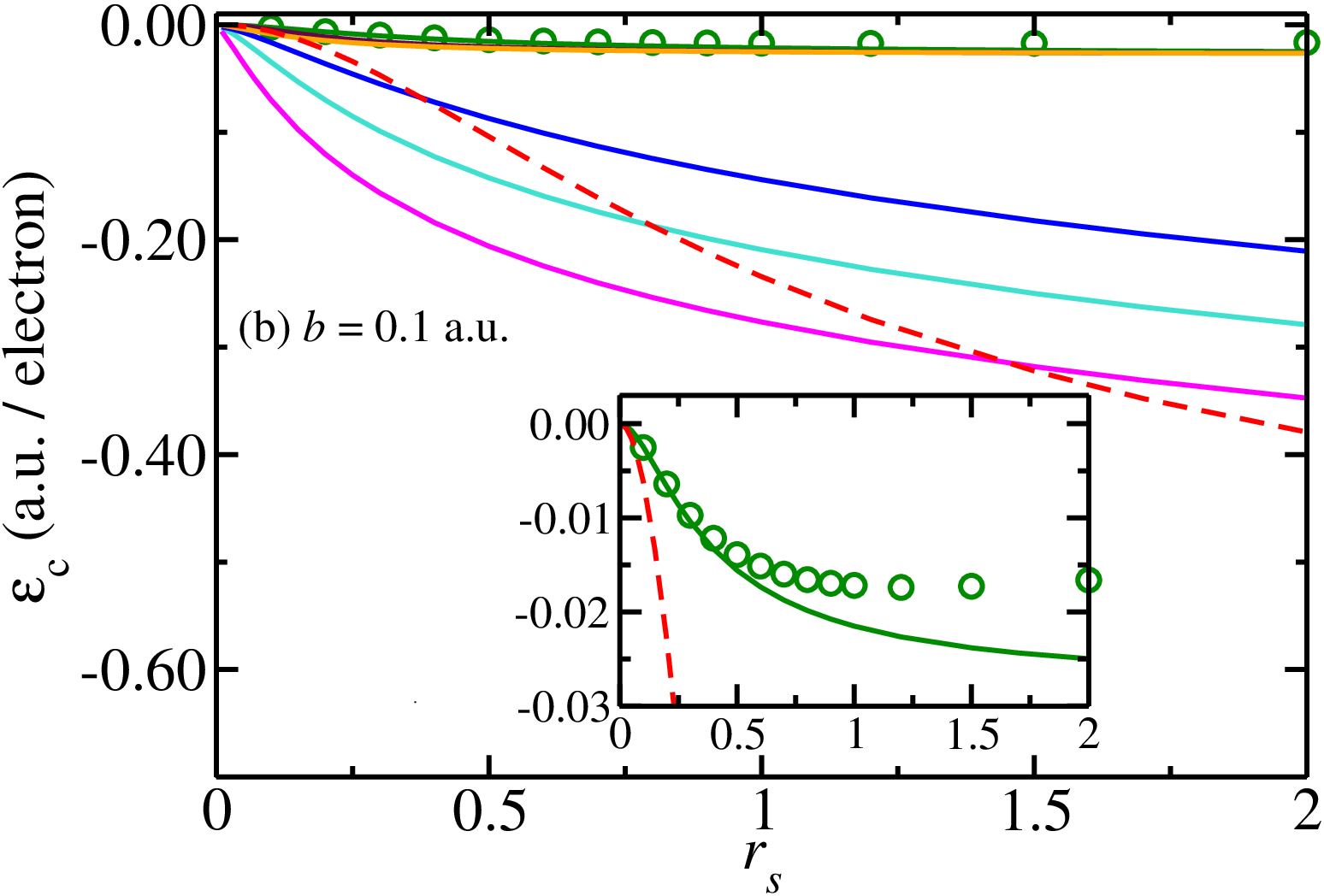}
\end{minipage}
\\[1em]
\begin{minipage}[t]{0.48\textwidth}
\includegraphics[clip,width=0.95\textwidth]{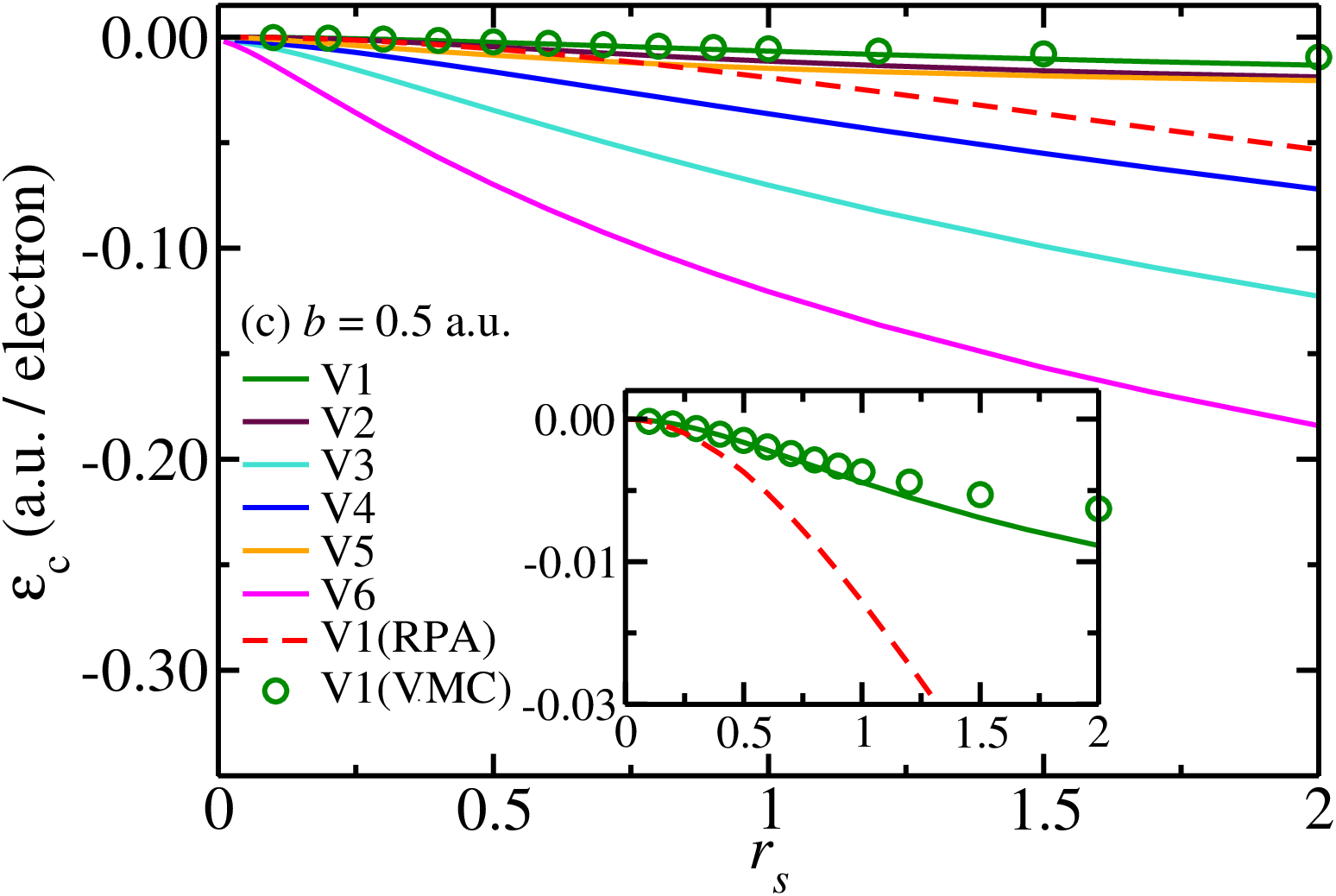}
\end{minipage}

\caption{Correlation energy per electron $\epsilon_\text{c}$ as a function of density parameter $r_\text{s}$ at fixed wire width $b=0.005$, $0.1$, and $0.5$ a.u.\ for different confinement models [$V_1(q)$--$V_6(q)$]. Inset of Fig.\ \ref{lce_vs_rs_bfixed}(a) depicts a zoomed in view of the correlation energy for $V_1(q)$,\, $V_2(q)$, and $V_5(q)$. Also, in the insets of Figs.\ \ref{lce_vs_rs_bfixed}(b) and \ref{lce_vs_rs_bfixed}(c), we have shown a zoomed-in view of the correlation energy and the validity of the FRPA with VMC simulation \cite{ashokan2018one,girdhar2022electron} and the RPA.}
\label{lce_vs_rs_bfixed}
\end{figure}

%----------------------------------------------------------------
\begin{figure}[htbp]
    \centering
    \hspace{-0.6cm}
    \includegraphics[clip,width=.48\textwidth]{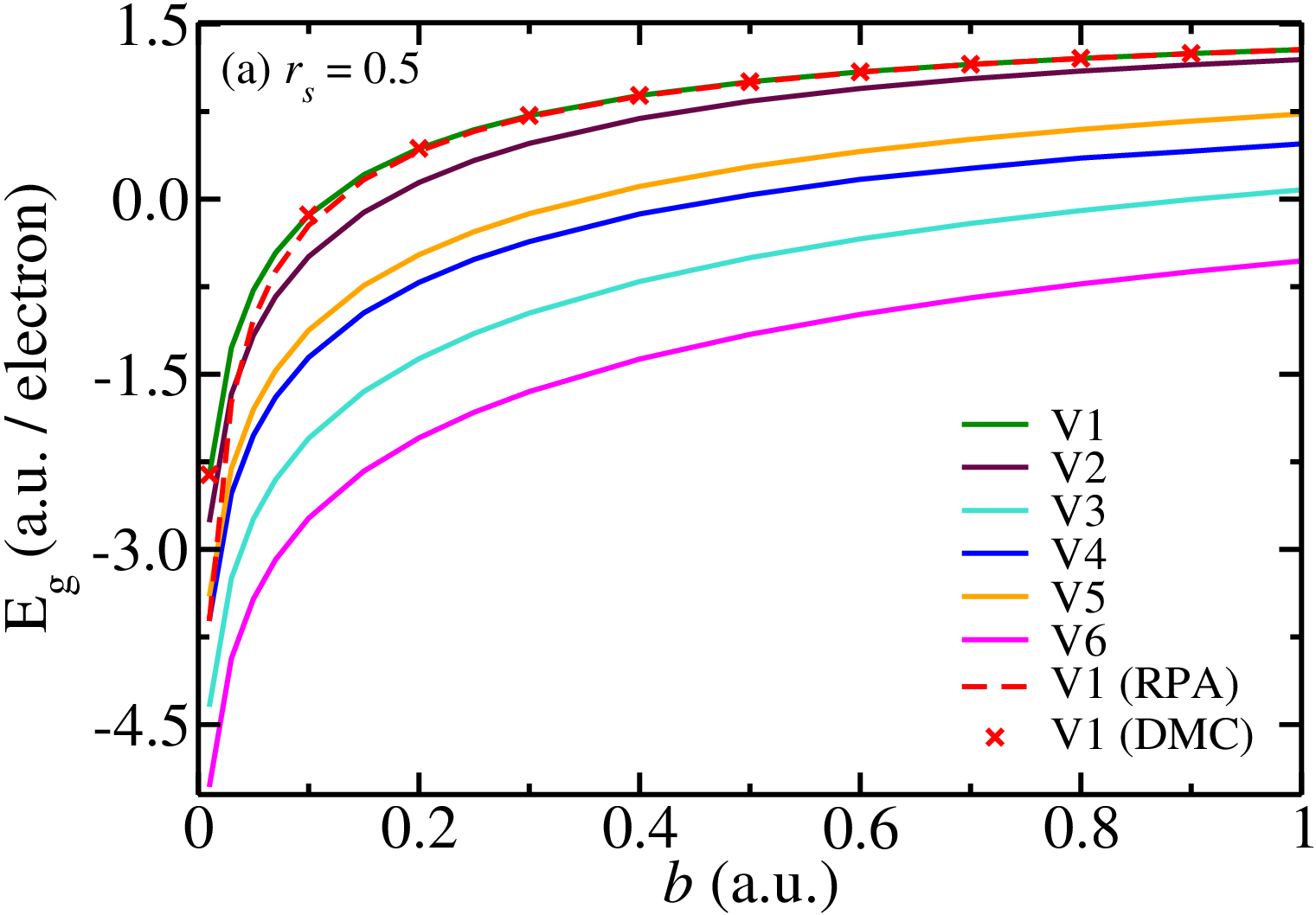}
    \vspace{0.4cm}
     \includegraphics[clip,width=.48\textwidth]{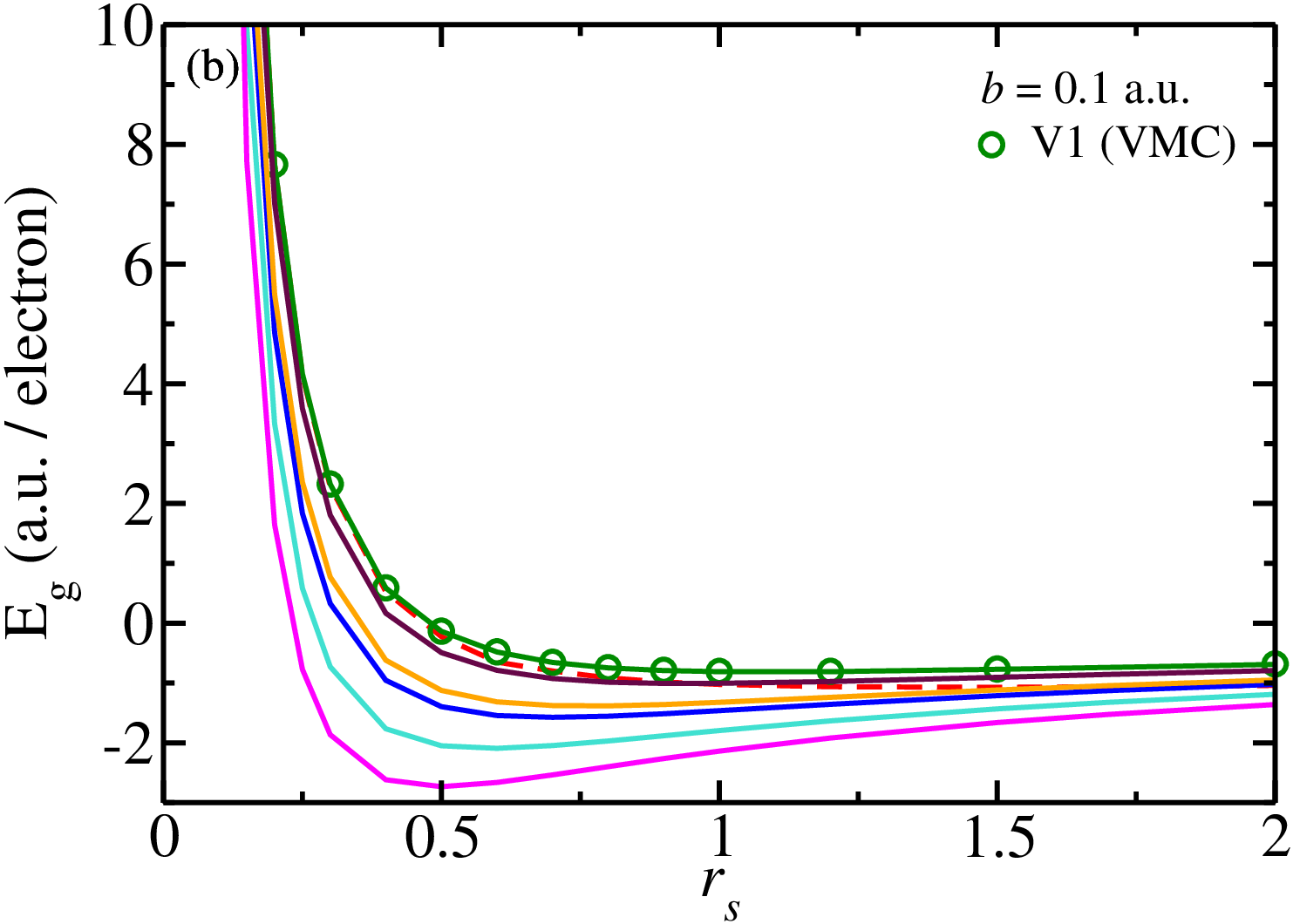}
    \vspace{0.4cm}
    \includegraphics[clip,width=.48\textwidth]{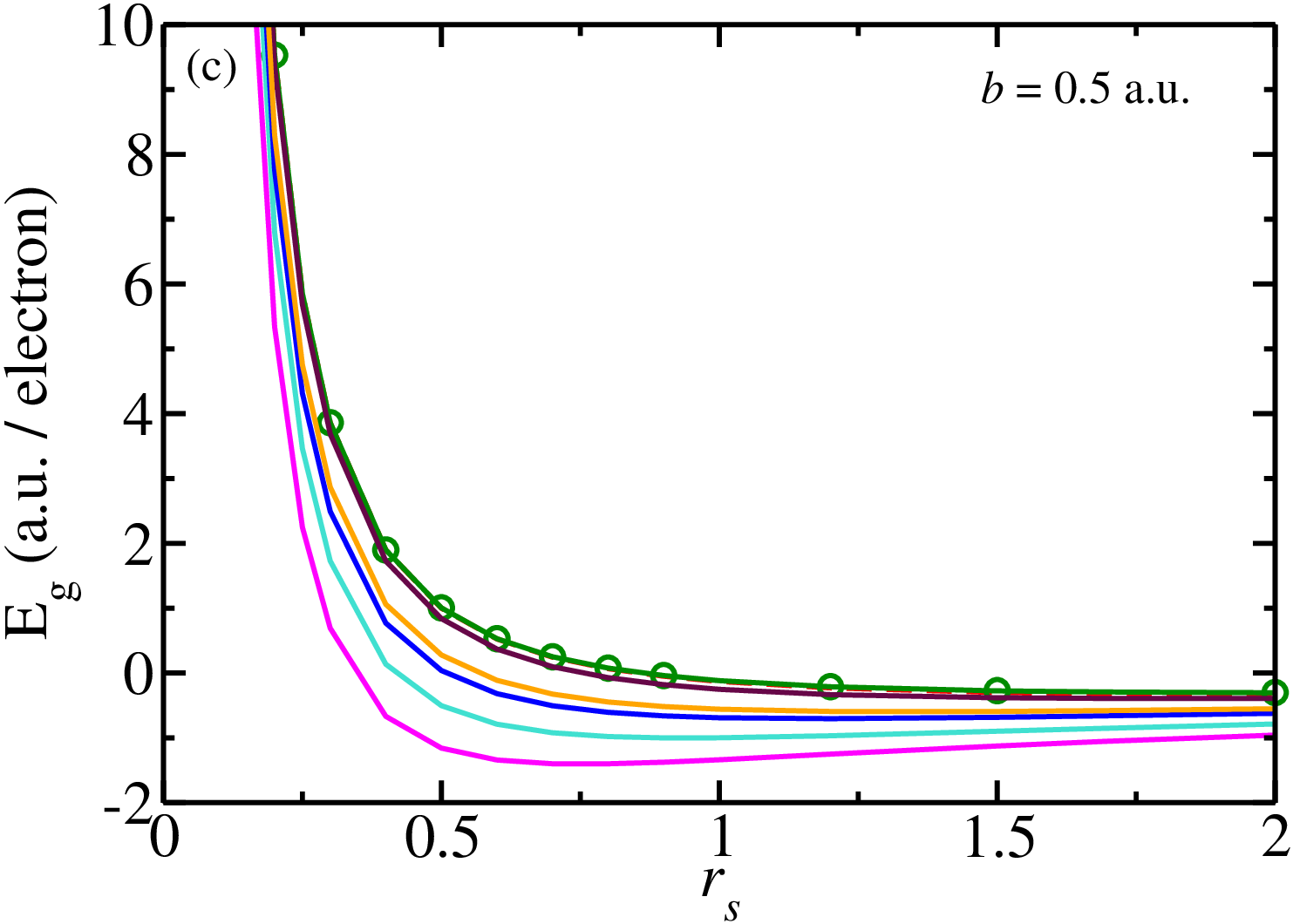}
    \caption{In Fig.\ \ref{lgse_vs_b_rs}(a) the ground state energy per electron $E_\text{g}$ is plotted as a function of wire width $b$ for $r_\text{s}=0.5$ for different confinement models. In Figs.\ \ref{lgse_vs_b_rs}(b) and \ref{lgse_vs_b_rs}(c), the ground state energy per electron $E_\text{g}$ is plotted as a function of density parameter $r_\text{s}$ for wire widths $b=0.1$\,\,\text{and}\,\, $0.5$ a.u. Also, the ground state energies in the FRPA are compared with QMC simulated and RPA results for harmonic wires \cite{ashokan2018one,girdhar2022electron}.}
    \label{lgse_vs_b_rs}
\end{figure}
%----------------------------------------------------------------

\section{Results and discussion}\label{results} In this section, we discuss the results of our numerical calculations of the SSF, PCF, correlation energy, and ground state energy of various confinement models as in Eqs.\ (\ref{V1}), (\ref{V2}), (\ref{V3}), (\ref{V4}), (\ref{V5}), and (\ref{V6}). 

\par The SSF is plotted for various density parameters and wire widths using Eq.\ (\ref{ssftotal}). In Fig.\ \ref{SSFtheoryrs_b} it is plotted for density parameters $r_\text{s} =0.6$, 0.8, 1.2, and $1.5$, with wire widths $b=0.1$ and $0.5$ a.u. 
The SSF shows a prominent peak at $2k_\text{F}$, of which a magnified view can be seen in the main plot of Fig.\ \ref{SSFtheoryrs_b}. Figures \ref{SSFtheoryrs_b}(a) and \ref{SSFtheoryrs_b}(b) show that, as $r_\text{s}$ varies from $0.6$ to $0.8$ at fixed $b=0.1$ a.u., the correlations between the electrons are enhanced and thus lead to higher values of the $2k_\text{F}$ peak heights in the SSF\@. Similarly, this can be observed in Figs.\ \ref{SSFtheoryrs_b}(c) and \ref{SSFtheoryrs_b}(d).
Further, it can be observed from Figs.\ \ref{SSFtheoryrs_b}(a) and \ref{SSFtheoryrs_b}(c) [also in Figs.\ \ref{SSFtheoryrs_b}(b) and \ref{SSFtheoryrs_b}(d)] that as $b$ increases the height of the $2k_\text{F}$ peak decreases at fixed $r_\text{s}$. The coupling between the electrons decreases and the system is less correlated for width $b=0.5$ a.u.\ compared to $b=0.1$ a.u.  For fixed $b=0.1$ and $0.5$ a.u., the trend of the peak heights at $2k_\text{F}$ in the SSF is linear with the density parameter $r_\text{s}$ ($b<r_\text{s}$), as shown in Figs.\ \ref{lssf_peaks_vs_rs_bfixed}(a) and \ref{lssf_peaks_vs_rs_bfixed}(b). We propose a fitting function for the parametric relationship of the peak height of the charge SSF $S(2k_\text{F},b)$ with wire width $b$ deduced from our finite width theory \cite{girdhar2022electron} and bosonization \cite{schulz1993wigner,fano1999unscreened}. 
The functional form  can be represented as 
\begin{eqnarray}\label{Eq:SSF2kf_fit}
    S(2k_\text{F},b) = \frac{a_0}{(b/r_\text{s})}\exp{\bigg(-c\sqrt{\ln{\frac{a_1\;b}{r_\text{s}}}}\bigg)} +  \;a_2,
\end{eqnarray} 
where $a_0$, $a_1$, $a_2$, and $c$ are fitting parameters.

Figures \ref{lssf_peaks_vs_b_rsfixed}(a) and \ref{lssf_peaks_vs_b_rsfixed}(b) show how the peak height of the SSF at $2k_\text{F}$ changes with wire-width $b$ for different confinement models at fixed $r_\text{s}$. The higher peaks generally suggest stronger electron correlations as $b$ decreases. Equation (\ref{Eq:SSF2kf_fit}) provides a good fit to the $2k_\text{F}$ peak heights, and the fitted results are consistent with our finite wire-width theory and bosonization.
Overall, Figs.\ \ref{SSFtheoryrs_b}, \ref{lssf_peaks_vs_rs_bfixed}, and \ref{lssf_peaks_vs_b_rsfixed} depict that the peak height is different for each confinement model and the strength of electron correlations is sensitive to how electrons are confined in the wire. We can conclude that the differences in peak heights between the models are more pronounced, as electron correlations are stronger for thinner wires and lower electron densities \cite{girdhar2023wire}. In Fig.\ \ref{SSFtheoryrs_b}, we have also compared our numerically calculated SSF with the VMC-simulated SSF for the confinement model [$V_1(q)$] of the harmonic wire \cite{ashokan2018one,girdhar2022electron}. 
We observe that the height of the $2k_\text{F}$ peak depends significantly on the density parameter $r_\text{s}$, wire width $b$, and confinement model $V_1(q)$--$V_6(q)$.
%--------------------------------------------------------------------------------------

We are studying a ferromagnetic system where pairs of electrons with parallel spins are affected by the antisymmetry of the wave function. For same-spin electrons, the PCF diminishes to zero (consistent with Pauli's exclusion principle) at short interelectronic distances and converges to one, leading to an uncorrelated system, at large distances \cite{Giuliani2008}. This can be seen in Fig.\ \ref{PCFtheory_rs_b}, which illustrates the $r/r_\text{s}$ dependence of the PCF $g(r)$ at fixed $r_\text{s}$ and $b$. In the main plot of Fig.\ \ref{PCFtheory_rs_b}, we can observe that the amplitude of oscillations for various confinement models reaches its maximum value around $r=2r_\text{s}$, signifying that the normalized probability of observing an electron pair with an interelectronic distance of $2r_\text{s}$ is the highest. These peaks are associated with an oscillation period of $2r_\text{s}$ in the PCF\@. For fixed $b$, Figs.\ \ref{PCFtheory_rs_b}(a) and \ref{PCFtheory_rs_b}(b) [also Figs.\ \ref{PCFtheory_rs_b}(c) and \ref{PCFtheory_rs_b}(d)] show that the strength of correlations enhances with increasing $r_\text{s}$, and the PCF exhibits oscillatory behavior, with the amplitude of oscillations growing with the density parameter $r_\text{s}$. For fixed $r_\text{s}$, intrawire correlation effects decrease with increasing $b$, and the amplitude of oscillations decays as observed from Figs.\ \ref{PCFtheory_rs_b}(a) and \ref{PCFtheory_rs_b}(c). This behavior is consistently observed in Figs.\ \ref{PCFtheory_rs_b}(b) and \ref{PCFtheory_rs_b}(d).
We also present a comparison between our numerically calculated PCF with VMC (in Fig.\ \ref{PCFtheory_rs_b})  and RPA [in Figs.\ \ref{PCFtheory_rs_b}(e) and \ref{PCFtheory_rs_b}(f)] for the harmonic wire confinement model \cite{ashokan2018one,girdhar2022electron}.

%--------------------------------------------------------------------------------------
The correlation energy is the difference between the exact ground state energy and the Hartree-Fock ground state energy. Using Eq.\ (\ref{eq:corr_e_analytic}), we numerically calculate the correlation energy per electron for different confinement models at a fixed value of the coupling parameters ($r_\text{s}$ or $b$) as a function of ($b$ or $r_\text{s}$ ) respectively. In Fig.\ \ref{lce_vs_b_rsfixed}, we plot the correlation energy as a function of the wire width $b$ at a fixed value of the density parameter ($r_\text{s} = 0.1 \,\,\text{and}\,\, 0.5$). In Fig.\ \ref{lce_vs_b_rsfixed}(a) for $r_\text{s}=0.1$, a significant difference in correlation energies among various confinement models is observed as $b$ increases, whereas as $b\rightarrow 0$ the correlation energy with confinement potentials $V_1(q)$, $V_2(q)$, and $V_5(q)$ approaches $-\pi^2/360$ \cite{Loos2013,ashokan2020exact}. In contrast the correlation energy with confinement potentials $V_3(q)$, $V_4(q)$, and $V_6(q)$ does not approaches this high density limit. At fixed $r_\text{s}$, a smaller $b$ implies that stronger confinement forces electrons closer together, enhancing the Coulomb interaction.  In Figs.\ \ref{lce_vs_b_rsfixed}(a) and \ref{lce_vs_b_rsfixed}(b), we have found a good agreement with QMC-simulated correlation energies of harmonic wires [$V_1(q)$] for $r_\text{s}=0.1$ and $0.5$.   
Figure \ref{lce_vs_rs_bfixed} depicts the $r_\text{s}$ dependence of the correlation energies across various confinement models at fixed wire widths ($b=0.005$, 0.1, and 0.5 a.u.), and we have also compared it with our QMC-simulated results \cite{ashokan2018one,girdhar2022electron}. In Fig.\ \ref{lce_vs_rs_bfixed}(b), they are in good agreement within the validity of high density theory $r_\text{s}\leq0.6$ and start deviating in the region $0.6\leq r_\text{s}\leq2$ at fixed $b=0.1$ a.u.  In Fig.\ \ref{lce_vs_rs_bfixed}(c), the FRPA is valid in range $r_\text{s}\leq1.2$ and shows deviation from QMC in the domain $1.2\leq r_\text{s}\leq2$ at $b=0.5$ a.u. Also we have shown the validity of RPA till $r_\text{s}\leq0.1$ for $b=0.1$ a.u.\ and $r_\text{s}\leq0.3$ for $b=0.5$ a.u. Unlike the confinements $V_1(q)$, $V_2(q)$, and $V_5(q)$, the correlation energies for the  confinement potentials $V_3(q)$, $V_4(q)$, and $V_6(q)$ are significantly different as $r_\text{s}$ increases. At fixed $b$, large $r_\text{s}$ implies a lower electron density; the electron-electron interactions become stronger and the magnitude of the correlation energy increases. In a high-density regime ($r_\text{s}<1$), the correlation energy is more sensitive to both wire width $b$ and choice of confinement model.This highlights that the choice of confinement model significantly affects the calculated correlation energy, especially at intermediate values of $b$. \par The ground-state energy is calculated by summing all the contributions outlined in Eq.\ (\ref{gse_all}). In Fig.\ \ref{lgse_vs_b_rs}, the ground-state energies for various confinement models are plotted against $b$ at $r_\text{s}=0.5$ and against $r_\text{s}$ at $b=0.1$ and 0.5 a.u. At fixed $r_\text{s} =0.5$, the ground-state energy per electron becomes negative at smaller wire width $b$ values for all confinement models as in Fig.\ \ref{lgse_vs_b_rs}(a). At fixed $b =0.1,\,0.5$ a.u., the ground-state energy per electron rapidly decreases and becomes negative. In simpler terms, the electrons arrange themselves in a way that minimizes their repulsive interactions, indicating that the interactions between the particles actually stabilize the system. It suggests that the 1D HEG becomes more energetically stable as the wire width $b$ decreases \cite{girdhar2022electron}. Obviously this is only because the transverse kinetic energy and the electrostatic energy of the background are neglected.

%--------------------------------------------------------------------------------------
%\vspace{-0.5cm} % Just for consistent with text alignment
\section{Conclusions}\label{Conclusion}
In this paper, we studied confinement-driven correlation effects on the ground-state properties of a ferromagnetic quasi-one-dimensional quantum wire using FRPA, which provides ground state structure beyond RPA, including self-energy and exchange contributions for various confinement models. The ground-state properties of quasi-one-dimensional quantum metallic wires for different confinement schemes have been calculated numerically and compared with the available QMC simulated data, which provides validation of the accuracy of our numerical calculations. Our study demonstrates that the choice of confinement model can have a significant impact on the ground-state properties of quasi-one-dimensional quantum wires. This suggests that employing an accurate confinement model when studying these systems becomes crucial. The prominent peaks of charge SSF at $2k_\text{F}$ are fitted using a fitting function based on our finite wire-width theory, guided by insights from bosonization, demonstrating good agreement with our FRPA theory.

Our findings also indicate that for the ultrathin wire in the high-density limit the correlation energy for confinement models $V_1(q)$, $V_2(q)$, and $V_5(q)$ approaches $\epsilon_\text{c}(r_\text{s})= - \pi^2/360$ a.u., which agrees with the exact results in this limit \cite{Loos2013, ashokan2020exact}. This clearly illustrates that for at least these three confinement potentials, the one-dimensional Coulomb potential can be regularized at interparticle distance $x=0$ to yield the same correlation energy. In contrast, other confinement potentials $V_3(q)$, $V_4(q)$, and $V_6(q)$ do not approach the same high-density limit. The percentage difference in correlation energy between the confinement models $V_1(q)$, $V_2(q)$, and $V_5(q)$ and $V_3(q)$, $V_4(q)$, and $V_6(q)$ is within about $10\%$  in the high-density limit. This indicates that the specific shape of the confinement potential affects the long-range correlations between the electrons. These findings contribute to a deeper understanding of electronic correlations and confinement effects in quasi-1D metallic systems. This enhanced modelling of electron interaction effects within confined systems may be useful in the optimization and fabrication of ferromagnetic quasi-1D quantum wire devices in the future.

\begin{acknowledgments}
One of the authors (V.G.)\ gratefully acknowledges the financial support from the Ministry of Education for carrying out this research work.
V.A.\ acknowledges the financial support of the Core Research Grant No.\ CRG/2023/001573 from the Science and Engineering Research Board, Anusandhan-National Research Foundation, Government of India. K.N.P.\ acknowledges the support received through the honorary scientist position awarded by the National Academy of Sciences, India Prayagraj.
We acknowledge National Supercomputing Mission (NSM) for providing computing resources of ``PARAM Siddhi-AI,'' under National PARAM
Supercomputing Facility (NPSF), C-DAC, Pune.
\end{acknowledgments}

\section*{data availability}

The data that support the findings of this article are openly available \cite{Vidit2025data}.

%----------------------------------------------------------------

\begin{widetext} 
 \section*{Appendix}
  In general, the Hamiltonian for a quasi-one-dimensional HEG can be written schematically as 
\begin{equation}\label{hamiltonian}
    \hat{H} = \sum_i \frac{\hat{p}_{i}^2}{2m} + \frac{1}{2}\sum_{i \neq j}\frac{e^2}{|{x_{i} - x_{j}}|} + \Lambda,
\end{equation}
where $\hat{p}_i$ is the momentum of the $i^{\text{th}}$ electron and the second term in Eq.\ (\ref{hamiltonian}) is the Coulomb interaction between electrons positioned at $r_i$ and $r_j$. $\Lambda$ represents Coulomb interactions with the background, required to make the electron-electron Coulomb sum convergent. The electron-electron Coulomb interaction is modified at short range due to different confinement potentials in the $yz$-plane.

\appendix
\section{Harmonic confinement}\label{AppendixA}  
With harmonic confinement, the electrons are free to move along the axis of the wire ($x$-direction) and are trapped in the ground state of a harmonic potential in the $yz$-plane. 
So, the wave function of one electron will have the following form:
\begin{equation}\label{twfhp}
    \psi_{q}(x,R) = \frac{1}{\sqrt{L}}\exp(iqx) \phi(R),
\end{equation}
where $R$ is a two-dimensional vector, $q$ is in the $x$-direction, and $\phi(R)$ is the two-dimensional harmonic oscillator wave function. Using Eqs.\ (\ref{hamiltonian}) and (\ref{twfhp}), the average electron-electron interaction is given by
\begin{equation}\label{average_eei}
V_1(q) = \int_{-\infty}^{\infty} e^{iq(x - x')} \, d(x - x')
\int \int \frac{|\phi(R)|^2 |\phi(R')|^2 \, dR \, dR'}
{\left[ (x - x')^2 + (R - R')^2 \right]^{1/2}}.
\end{equation}
The electron transverse wave function for two-dimensional harmonic oscillators in cylindrical coordinates is given as 
\begin{equation}
    \phi(R) = \left( 2 \pi b^2 \right)^{-1/2} \exp\left( - \frac{R^2}{4b^2} \right),
\end{equation}
where $b$ is related to wire width and oscillator frequency $\omega_0 = \frac{\hbar}{2m^*b^2}$. Now, substituting electron wave function for two-dimensional harmonic oscillators in Eq.\ (\ref{average_eei}), we get
\begin{equation}
    V_1(q)= \int_{-\infty}^{\infty} e^{i q x} \, d x\left(\frac{1}{2 \pi b^{2}}\right)^{2} \int \int \frac{\exp \left(\frac{-R^{2}}{2 b^{2}}\right) \exp \left(\frac{-R^{\prime 2}}{2 b^{2}}\right)}{\left[x^{2}+\left(R-R^{\prime}\right)^{2}\right]^{1 / 2}} \, d R \, d R^{\prime}.
\end{equation}
Using $\alpha=\frac{1}{2 b^{2}}$ and substituting $R-R^{\prime}=R^{\prime \prime}$ we obtain,
\begin{align}
    V_1(q)&=\frac{1}{4  \pi^{2} b^{4}} \int_{-\infty}^{\infty} e^{i q x} \, d x \int \int \frac{\exp \left(-\alpha R^{2}\right) \exp \left(-\alpha\left(R-R^{\prime \prime}\right)^{2}\right)}{\left[x^{2}+R^{\prime \prime 2}\right]^{1 / 2}} \, d R \, d R^{\prime \prime}\\
    V_1(q)& =\frac{1}{4 \pi^{2} b^{4}} \int_{-\infty}^{\infty} e^{i q x} \, d x \int \int \frac{e^{-2\alpha R^{2}} e^{-\alpha R^{\prime 2}} e^{2 \alpha R R^{\prime} \cos \theta} R R^{\prime} \, d R \, d R^{\prime} \, d \theta \, d \theta^{\prime}}{\left[x^{2}+R^{\prime 2}\right]^{1 / 2}},\\ 
    &=\frac{1}{2 \pi b^{4}} 2 \pi \int_{-\infty}^{\infty} e^{i q x} \, d x \int \int \frac{e^{-2 \alpha R^{2}} e^{-\alpha R^{\prime 2}} R R^{\prime} \, d R \, d R^{\prime}}{\left[x^{2}+R^{\prime 2}\right]^{1 / 2}}\left(\frac{1}{\pi} \int_{0}^{\pi} e^{2 \alpha R R^{\prime} \cos \theta} \, d \theta\right).
\end{align}
Using angular integration and the zeroth-order Bessel's function of the first kind, $J_{0}(z) = \frac{1}{\pi}\int_{0}^{\pi}e^{iz\cos{\theta}}\,d\theta$,
\begin{equation}\label{10}
    V_1(q)=\frac{1}{ b^{4}} \int_{-\infty}^{\infty} e^{i q x} \, d x \int  \frac{ e^{-\alpha R^{\prime 2}} R R^{\prime} \, d R^{\prime}}{\left[x^{2}+R^{\prime 2}\right]^{1 / 2}} \int_{0}^{\infty} e^{-2 \alpha R^{2}} R J_{0}\left(2 \alpha R R^{\prime}i\right)
    \, d R.
\end{equation}
Integration over $R$ in Eq.\ (\ref{10}) gives,
\begin{equation}\label{11}
   V_1(q)=\frac{1}{4 \alpha b^{4}} \int_{-\infty}^{\infty} e^{i q x} \, d x \int_{0}^{\infty} \frac{e^{-\frac{\alpha R^{\prime 2}}{2} } R^{\prime} \, d R^{\prime}}{\left[x^{2}+R^{\prime 2}\right]^{1 / 2}}.
\end{equation}

Now substituting $R^{\prime 2}=t$ in Eq.\ (\ref{11}) reduces to
\begin{equation}\label{12}
 V_1(q)=\frac{1}{4 b^{2}} \int_{-\infty}^{\infty} e^{i q x} \, d x \int_{0}^{\infty} \frac{e^{-\frac{\alpha}{2} t} \, d t}{\left[x^{2}+t\right]^{1 / 2}}. \end{equation}
Again substituting, $x^{2}+t=u^{2}$. As $t  =0,\hspace{0.1cm} u=|x| \hspace{0.2cm} \text{and} \hspace{0.2cm}   t \rightarrow \infty,\hspace{0.1cm} u \rightarrow \infty$. Equation (\ref{12}) can be written as
\begin{align*}
 V_1(q) & =\frac{1}{4 b^{2}} \int_{-\infty}^{\infty} e^{i q x} \, d x \int_{|x|}^{\infty} e^{-\frac{\alpha}{2}\left(u^{2}-x^{2}\right)} \, d u, \\
& =\frac{1}{4 b^{2} } \int_{-\infty}^{\infty} e^{i q x} e^{\frac{\alpha}{2} x^{2}} \, d x\left[\int_{|x|}^{0} e^{-\frac{\alpha}{2} u^{2}} \, d u+\int_{0}^{\infty} e^{-\frac{\alpha}{2} u^{2}} \, d u\right].
\end{align*}
Now changing the variable from $u$ to $y$ by relation, $\frac{\alpha}{2} u^{2}=y^2$,
which implies the limit of integration as, $u=|x| \rightarrow y=\sqrt{\frac{\alpha}{2}}|x|$. Therefore it becomes
\begin{equation}\label{14}
   V_1(q)=\frac{1}{2 b^{2}} \int_{-\infty}^{\infty} e^{i q x} e^{\frac{\alpha}{2} x^{2}} \, d x \sqrt{\frac{2}{\alpha}}\left[ \int_{\frac{|x|}{2 b}}^{0} e^{-y^{2}} \, d y+\int_{0}^{\infty} e^{-y^{2}} \, d y\right].
\end{equation}
We can rewrite Eq.\ (\ref{14}) as
\begin{equation}
 V_1(q)=\frac{\sqrt{\pi}}{2b} \int_{-\infty}^{\infty} e^{i q x} e^{\frac{x^{2}}{4 b^{2}}} \operatorname{erfc}\left(\frac{|x|}{2 b}\right) \, d x,  
 \end{equation}
where  $v_1(x)=\frac{\sqrt{\pi}}{2b} \exp \left(\frac{x^{2}}{4 b^{2}}\right) \operatorname{erfc}\left(\frac{|x|}{2 b}\right)$ in which erfc is the complementary error function defined as $\operatorname{erfc}(z) = \frac{2}{\sqrt{\pi}}\int_{z}^{\infty} e^{-t^2} \, dt$. It can be noted that $V(x)$ is finite at $x=0$ and varies as $1 / x$ for $x \rightarrow \infty$. Hence
\begin{equation}\label{16}
    V_1(q)= \frac{\sqrt{\pi}}{2b} \int_{-\infty}^{\infty} d x \, e^{i q x} e^{\frac{x^{2}}{4 b^{2}}}\left(1-\varphi\left(\frac{x}{2 b}\right)\right),
\end{equation}
where $\varphi\left(\frac{x}{2 b}\right)$ is the error function. The integral in Eq.\ (\ref{16}) can be simplified by using the method of contour integration by choosing a closed contour in the first quadrant of the upper half complex $x$ plane and noting that the integral is zero, as there is no pole in the integrand and the contribution of the quarter circle also vanishes when the magnitude of complex $x \rightarrow \infty$. This procedure converts the integral along the real axis to an integral along the imaginary $y$-axis, which reduces the integral to a standard form that can be easily integrated, and Eq.\ (\ref{16}) becomes:
\begin{equation}
V_1(q)= \frac{\sqrt{\pi}}{b}\left\{\frac{-b}{\sqrt{\pi}} e^{q^{2} b^{2}} E_{1}\left(-q^{2} b^{2}\right)-i \sqrt{\pi} b e^{a^{2} b^{2}}\right\}.
\end{equation}
This can be written as
\begin{align}
V_1(q) & = e^{q^{2} b^{2}}\left\{-E_{1}\left(-q^{2} b^{2}-i 0\right)-i \pi\right\}  \\
& = e^{q^{2} b^{2}}\left\{E_{1}\left(q^{2} b^{2}\right)+i \pi-i \pi\right\},
\end{align}
which can be written as
\begin{equation}
    V_1(q)= \exp \left(q^{2} b^{2}\right) E_{1}\left(q^{2} b^{2}\right).
\end{equation}
The limiting behavior of harmonic confinement is
\begin{align}
        V_1(q)& = -  \gamma -2\ln(q\,b) \quad \text{for} \quad (q\,b) \rightarrow 0,\\
        & =\frac{1}{(q\,b)^{2}} \,\,\quad\quad\quad\quad \text{for}\quad (q\,b) \rightarrow \infty,\label{V1atlargeq}
\end{align}
where $\gamma$ is the Euler constant.
% -------------------------------------------------------------------------------------------------
\section{Cylindrical confinement} \label{AppendixB}
\subsection{Radial Distribution of Cylindrically Confined Electrons}

Consider an electron confined in a cylindrical channel of radius $b$, with hard-wall boundary conditions on the cylindrical channel. Suppose $b$ is small compared to the mean electron separation. Then the transverse motion is dominated by the kinetic energy of confinement, and hence we can describe the transverse motion in an independent particle approximation.

In cylindrical polar coordinates, the wave function of a single electron in the channel is of the form 
\begin{eqnarray}
 \Psi({\bf r}) = R(r, \theta) X(x),
\end{eqnarray}
where $r$ and $\theta$ are the radial and polar coordinates, and $x$ is the axial coordinate. In the ground state, $R$ is independent of $\theta$. The radial Schr\"{o}dinger equation for the ground state is
\begin{eqnarray}
  - \frac{1}{2} \left( R^{\prime\prime}(r) + \frac{1}{r} R^{\prime}(r) \right) = E_R R(r),
\end{eqnarray}
where $E_R$ is the constant contribution to the energy eigenvalue due to the kinetic energy of the transverse motion. $R(r)$ is smooth and nondivergent for $0 \leq r < b$ and satisfies the boundary condition $R(b) = 0$.
Hence we have
\begin{eqnarray}
    r^2 R^{\prime\prime}(r) + r R^{\prime}(r) + 2E_R r^2 R(r) = 0.
\end{eqnarray}

Let $r = \alpha s$ for some constant $\alpha$. Let $S(s) = R(r)$. Then $S^{\prime}(s) = \alpha R^{\prime}(r) \quad \text{and} \quad S^{\prime\prime}(s) = \alpha^2 R^{\prime\prime}(r)$. Hence, the equation becomes
\begin{eqnarray}
 s^2 S^{\prime\prime}(s) +  s S^{\prime}(s) + 2 E_R \alpha^2 s^2 S(s) = 0.
\end{eqnarray}
But $E_R$ is positive since it is a purely kinetic energy. Let $\alpha = 1/\sqrt{2 E_R}$. Then
\begin{eqnarray}
 s^2 S^{\prime\prime}(s) + s S^{\prime}(s) + s^2 S(s) = 0.
\end{eqnarray}
This is Bessel's equation of zeroth order. The nondivergent solution is $S(s) = J_0(s)$, i.e., 
\begin{eqnarray}
R(r) = J_0\left( r/\alpha \right) = J_0\left( \sqrt{2 E_R} r \right),
\end{eqnarray}
where $J_0$ is the zeroth order Bessel function of the first kind. 

In the ground state, $R(b) = 0$, and there are no additional nodes in $0 \leq r < b$. Hence, 
$\sqrt{2 E_R} b = j_{0,1}$, where $j_{0,1} \approx 2.40483 \dots$ is the first zero of $J_0$. So the radial wave function is $R(r) = J_0\left( j_{0,1} \,r/b \right)$  in $0 \leq r \leq b$.
Hence the radial distribution is
\begin{eqnarray}
\left| \tilde{R}(r) \right|^2 \equiv \frac{|R(r)|^2}{\int_0^b 2\pi r |R(r)|^2 \, dr}.
\end{eqnarray}

\subsection{Effective Interaction between Cylindrically Confined Electrons}

Now consider two cylindrically confined electrons (labelled 1 and 2) with axial separation $x$. Assuming tight confinement in the transverse direction, we average the full Coulomb interaction over the radial distributions to obtain the effective Coulomb interaction
\begin{eqnarray}
v_2^{\text{eff}}(x) = \int_0^b \int_0^{2\pi} \int_0^b \int_0^{2\pi} \left| \tilde{R}(r_1) \right|^2 \left| \tilde{R}(r_2) \right|^2 \frac{1}{|{\bf r}_2 - {\bf r}_1|}\, r_1 \, d\theta_1 \, dr_1 \, r_2 \, d\theta_2 \, dr_2.
\end{eqnarray}

But ${\bf r}_2 - {\bf r}_1 = x \hat{\bf e}_x + \left[ r_2 \cos(\theta_2) - r_1 \cos(\theta_1) \right] \hat{\bf e}_y + \left[ r_2 \sin(\theta_2) - r_1 \sin(\theta_1) \right] \hat{\bf e}_z $, where $\hat{\bf e}_\beta$ is the unit vector in the Cartesian direction $\beta$. Hence
\begin{align}
|{\bf r}_2 - {\bf r}_1|^2 & =  x^2 + r_2^2 \cos^2(\theta_2) + r_1^2 \cos^2(\theta_1) - 2r_1r_2 \cos(\theta_1) \cos(\theta_2) + r_2^2 \sin^2(\theta_2) + r_1^2 \sin^2(\theta_1) - 2r_1r_2 \sin(\theta_1) \sin(\theta_2) \nonumber \\
 & =x^2 + r_1^2 + r_2^2 - 2r_1r_2 \cos(\theta_2 - \theta_1).
\end{align}
So,
\begin{eqnarray}\label{Veff_int1}
  v_2^{\text{eff}}(x) = 2\pi \int_0^b \int_0^b \int_0^{2\pi} r_1 r_2 \left| \tilde{R}(r_1) \right|^2 \left| \tilde{R}(r_2) \right|^2 \frac{1}{\sqrt{x^2 + r_1^2 + r_2^2 - 2r_1r_2 \cos(\theta)}} \, d\theta \, dr_2 \, dr_1.
\end{eqnarray}

To find the effective interaction, we must solve the integral in Eq.\ (\ref{Veff_int1}). We can make a little bit of progress by solving the angular integral:
\begin{align}
\int_0^{2\pi} \frac{d\theta}{\sqrt{x^2 + r_1^2 + r_2^2 - 2r_1r_2 \cos(\theta)}} 
& = \int_0^{2\pi} \frac{d\theta}{\sqrt{x^2 + r_1^2 + r_2^2 - 2r_1r_2 + 4r_1r_2 \sin^2(\theta/2)}} \nonumber \\
& =4\int_0^{\pi/2} \frac{d\theta}{\sqrt{x^2 + (r_1 - r_2)^2 + 4r_1r_2 \sin^2(\theta)}} \nonumber \\
& = \frac{4} {\sqrt{x^2 + (r_1 - r_2)^2} }\int_0^{\pi/2} \frac{d\theta}{\sqrt{1 + \frac{4r_1r_2}{x^2 + (r_1 - r_2)^2} \sin^2(\theta)}} \nonumber \\
& = \frac{4} {\sqrt{x^2 + (r_1 - r_2)^2} }K\left(k^2 = -\frac{4r_1r_2}{x^2 + (r_1 - r_2)^2}\right),
\end{align}
where $K$ is a complete elliptic integral of the first kind. Substituting this in Eq.\ (\ref{Veff_int1}) gives, 
\begin{eqnarray}\label{Veff_cyl_int}
v_2^{\text{eff}}(x) = 8\pi \int_0^b \int_0^b r_1r_2 \left| \tilde{R}(r_1) \right|^2 \left| \tilde{R}(r_2) \right|^2 \frac{K\left(k^2 = -\frac{4r_1r_2}{x^2 + (r_1 - r_2)^2}\right)}{\sqrt{x^2 + (r_1 - r_2)^2}} \, dr_1 \, dr_2.
\end{eqnarray}
As a sanity check, in the limit of large $x$, $\sqrt{x^2 + (r_1 - r_2)^2} \rightarrow |x|$ and $K\left(k^2 = -\frac{4r_1r_2}{x^2 + (r_1 - r_2)^2}\right) \rightarrow K(0) = \pi/2$.

Hence, at long range, the effective interaction reduces to the bare interaction:
\begin{eqnarray}\label{Veff_cyl_int_zero}
v_2^{\text{eff}}(x) \rightarrow \frac{1}{|x|} \left( 2\pi \int_0^b r \left| \tilde{R}(r) \right|^2 \, dr \right)^2 = \frac{1}{|x|}.
\end{eqnarray}
The above Eq.\ (\ref{Veff_cyl_int_zero}) is for zero radius cylinder. It is difficult to make further analytic progress with Eq.\ (\ref{Veff_cyl_int}) when $\tilde{R}(r)$ is nontrivial, as is the case for cylindrical hard-wall confinement.

The above potential can be analyzed in an alternative way using the one-electron wave function of the form 
\begin{equation}
  \psi(x, y, z)=\frac{{e}^{iqx}}{\sqrt{L}} \sqrt{\delta(y)} \sqrt{\delta(z)},
\end{equation}
where $\sqrt{\delta(y)}$ and  $\sqrt{\delta(z)}$ are types of wave functions which are equivalent to the assumption of zero wire width in the $y$ and $z$-directions, resulting in a much harder interaction potential.

Matrix elements of the Coulomb interaction are given by
\begin{eqnarray}
    V_2(q) &=&\int_{-\infty}^{\infty}  e^{iq(x-x^\prime)} \, d(x-x^\prime)\int_{-\infty}^{\infty}\int_{-\infty}^{\infty}\int_{-\infty}^{\infty}\int_{-\infty}^{\infty} \frac{\delta(y)\delta{(y^\prime)} {\delta(z)\delta(z^\prime)} \, dy \, dy^\prime \, dz \, dz^\prime }{{\left[(x-x^\prime)^{2}+({y-y^\prime+b_1})^{2}+(z-z^\prime+b_2)^{2}\right]^{1 / 2}}}.
\end{eqnarray}
Substituting $x-x^\prime=x$ and using the Dirac delta function's integral,
\begin{equation}
    V_2(q) = \int_{-\infty}^{\infty} e^{iqx} \frac{1}{\left[x^{2}+b^{2}\right]^{1 / 2}} dx \,\,\,\,\,\,\,\,;\,\,\,\,\,\, b^2={b_1}^2+{b_2}^2,
\end{equation}
where $b$ is wire width parameter. 
Hence, the softened  Coulomb interaction is approximated as $v_2^{\text{eff}}(x)\approx 1/\sqrt{(x^2+b^2)}$ and Fourier transform after simplification becomes
\begin{equation}\label{Veff_fourier_int_cyl}
    V_2(q) = 2 \int_{0}^{\infty} \frac{\cos{(qx)}}{\left[x^{2}+ b^2\right]^{1 / 2}} \, dx.
\end{equation}

Using the integral identity for Bessel functions, Eq.\ (\ref{Veff_fourier_int_cyl}) becomes
\begin{equation}
    V_2(q) = 2K_{0}(qb),
\end{equation}
where $K_{0}(q b)$ is the zeroth-order modified Bessel function of the second kind.
The limiting behavior of cylindrical confinement is
\begin{align}
        V_2(q)& =  -2\gamma +2\ln2 -2\ln(q\,b) \quad \text{for} \quad (q\,b) \rightarrow 0, \\
        & = e^{-qb} \sqrt{\frac{2\pi}{q\,b}} \,\quad\quad\quad\quad\quad\quad\,\,\,\text{for}\quad (q\,b) \rightarrow \infty.\label{V2atlargeq}
\end{align}
% -------------------------------------------------------------------------------------------------

\section{Infinite square well confinement}\label{AppendixC}
For this case, the one-electron wave function is of the form
\begin{equation}\label{twf3}
 \psi(x, y, z)=\frac{{e}^{iqx}}{\sqrt{L}} \phi_{n}(y) \phi_{n}^{\prime}(z).
\end{equation}
 The ground state wave function for an infinite square well is
  \begin{eqnarray}\label{wf31}
      \phi(y) & = & \left(\frac{2}{b}\right)^{1 / 2} \sin \left(\frac{\pi y}{b}\right),\quad 0 \leq y \leq b \\
      \label{wf32}
      \phi^{\prime}(z) & = & \left(\frac{2}{b}\right)^{1 / 2} \sin \left( \frac{\pi z}{b}\right),\quad 0 \leq z \leq b.
 \end{eqnarray}
 Using Eqs.\ (\ref{wf31}) and (\ref{wf32}), the total wave function (\ref{twf3}) can be written as
 \begin{equation}
      \psi(x, y, z)=\frac{e^{i q x}}{\sqrt{L}}\left(\frac{2}{b}\right)^{1 / 2}\sin \left( \frac{\pi y}{b}\right)\left(\frac{2}{b}\right)^{1 / 2} \sin \left( \frac{\pi z}{b}\right),
 \end{equation}
where $b$ is the width of the infinite square well. Now, the matrix elements of the Coulomb interaction are given by
\begin{eqnarray}   
  V_3(q)&=& \frac{16}{b^{4}}   \int_{0}^{b} d z \int_{0}^{b} d z^{\prime} \, \sin ^{2}\left(\frac{\pi z}{b} \right)\sin ^{2}\left(\frac{\pi z^{\prime}}{b} \right)\int_{0}^{b} d y \int_{0}^{b} d y^{\prime} \, \sin ^{2}\left(\frac{\pi y}{b} \right) \sin ^{2}\left(\frac{\pi y^{\prime}}{b}\right)\nonumber\\
  &\times &\int_{-\infty}^{\infty} \frac{e^{i q(x-x^{\prime})}\, d(x-x^{\prime})}{\sqrt{(x-x^{\prime})^{2}+\left(y-y^{\prime}\right)^{2}+\left(z-z^{\prime}\right)^{2}}},
\end{eqnarray}
Substituting $x-x^\prime=x$,
\begin{eqnarray}\label{CIv3}
 V_3(q)&=& \frac{16}{b^{4}}  \int_{0}^{b} d z \int_{0}^{b} d z^{\prime} \, \sin ^{2}\left(\frac{\pi z}{b} \right)\sin ^{2}\left(\frac{\pi z^{\prime}}{b} \right)\int_{0}^{b} d y \int_{0}^{b} d y^{\prime} \, \sin ^{2}\left(\frac{\pi y}{b} \right) \sin ^{2}\left(\frac{\pi y^{\prime}}{b} \right)\nonumber\\ &\times& \int_{-\infty}^{\infty} \frac{e^{i qx} \, d x}{\sqrt{x^{2}+\left(y-y^{\prime}\right)^{2}+\left(z-z^{\prime}\right)^{2}}}.
\end{eqnarray}
Using the integral identity for Bessel functions, Eq.\ (\ref{CIv3}) becomes
\begin{equation}
   V_3(q)= \frac{32}{b^{4}}  \int_{0}^{b} d z \int_{0}^{b}  d z^{\prime} \, \sin ^{2}\left(\frac{\pi z}{b}\right) \sin ^{2}\left(\frac{\pi z^{\prime}}{b}\right)  \int_{0}^{b} d y\int_{0}^{b}  d y^{\prime} \, \sin ^{2}\left(\frac{\pi y}{b}\right) \sin ^{2}\left(\frac{\pi y^{\prime}}{b} \right) K_{0}(q R),
\end{equation}
where $R=\sqrt{\left(y-y^{\prime}\right)^{2}+\left(z-z^{\prime}\right)^{2}}$.

% -------------------------------------------------------------------------------------------------
\section{Infinite square well confinement in the $y$ direction, triangular quantum well confinement in the $z$ direction}\label{AppendixD}
The one-electron electron wave function for this case is given as
\begin{equation}\label{twf4}
   \psi(x, y, z)=\frac{e^{iqx}}{\sqrt{L}}  \phi_{n}(y) \zeta_{i}(z).
\end{equation}
The ground state wave function for an infinite square well in the $y$-direction is
\begin{equation}\label{wf41}
   \phi_(y)=\left(\frac{2}{b}\right)^{1 / 2} \sin \left(\frac{ \pi y}{b}\right), \quad 0 \leq y \leq b,
\end{equation}
and the ground state normalized wave function in the $z$-direction is
 \begin{equation}\label{wf42}
   \zeta(z)=\frac{1}{\sqrt{2}}\left(\frac{3}{z_0 }\right)^{3 / 2} z  \exp{\left(\frac{-3\,z }{2\,z_0}\right)} \Theta(z),
\end{equation}
where this $b$ is the width of 1-D wire along $y$ axis, and $z_0 $  is
 defined as the average wire width along the $z$-direction. Using Eqs.\ (\ref{wf41}) and (\ref{wf42}), the total wave function (\ref{twf4}) can be written as
\begin{equation}
   \psi(x, y, z)=\frac{e^{i k x}}{\sqrt{L}} \left(\frac{2}{b}\right)^{1 / 2} \sin \left(\frac{\pi y}{b}\right) \frac{1}{\sqrt{2}}\left(\frac{3}{{z_0}}\right)^{3 / 2}  z \exp{\left(\frac{-3\,z }{2\,z_0}\right)} \Theta(z). 
\end{equation}
Matrix elements of the Coulomb interaction are given by
\begin{eqnarray}
    \label{CIv4}
 V_4(q)&=&  \frac{729}{b^{2}\,{z_0}^6}  \int_{0}^{\infty} d z \int_{0}^{\infty} d z^{\prime}\, z^2 z^{\prime2} e^{-\frac{3\, (z + z^{\prime})}{z_0}}\int_{0}^{b} d y \int_{0}^{b} d y^{\prime} \, \sin^2\left(\frac{\pi y}{b}\right) \sin ^{2}\left(\frac{\pi y^{\prime}}{b}\right) \nonumber\\
 &\times& \int_{-\infty}^{\infty} \frac{e^{i q (x-x^{\prime})} \, d(x-x^{\prime})}{\sqrt{(x-x^{\prime})^{2}+\left(y-y^{\prime}\right)^{2}+\left(z-z^{\prime}\right)^{2}}},
\end{eqnarray}
using the integral identity for Bessel functions, Eq.\ (\ref{CIv4}) becomes
\begin{equation}
  V_4(q)=\frac{1458}{b^{2}\,{z_0}^6}
  \int_{0}^{\infty} d z \int_{0}^{\infty} d z^{\prime} \, {z^{2}} \,{z^{\prime 2}} \,e^{-\frac{3\, (z + z^{\prime})}{z_0}}\int_{0}^{b} d y \int_{0}^{b} d y^{\prime} \, \sin^2\left(\frac{\pi y}{b}\right) \sin ^{2}\left(\frac{\pi y^{\prime}}{b}\right)\,K_{0}(q R).
\end{equation}

% -------------------------------------------------------------------------------------------------
\section{Harmonic confinement in the $y$ direction, delta-function confinement in the $z$ direction}\label{AppendixE} 
In this case, we confine electrons in the transverse $yz$-plane using harmonic confinement in the $y$-direction and neglect the motion of electrons in the $z$-direction, which is equivalent to the assumption of zero wire width. 
The one-electron wave function is of the form
\begin{equation}\label{twf5}
\psi(x, y, z)=\frac{e^{iqx}}{\sqrt{L}} \xi_{n}(y)\,\sqrt{\delta(z)}.
\end{equation}
The ground state wave function for a harmonic potential in the $y$-direction is
\begin{equation}\label{wf51}
\xi_{0}(y)=\left[\frac{1}{\sqrt{\pi b}}\right]^{1 / 2} \exp\left({-\frac{y^{2}}{2 b^{2}}}\right) H_{0}\left(\frac{y}{b}\right),
\end{equation}
where $b$ is related to wire width and $H_{0}\left(y/{b}\right)=1$ is the Hermite polynomial. Using Eqs.\ (\ref{twf5}) and (\ref{wf51}), matrix elements of the Coulomb interaction are given by,
\begin{align}
  V_5(q)=& \int_{-\infty}^{\infty} e^{i q\left(x-x^{\prime}\right)} \, d\left(x-x^{\prime}\right)\int_{-\infty}^{\infty} dz^\prime\int_{-\infty}^{\infty}{\delta(z)\delta(z^\prime) \,dz} \int_{-\infty}^{\infty} d y \int_{-\infty}^{\infty} d y^{\prime} \frac{\xi_{0}{ }^{2}\left(y\right) \xi_{0}{ }^{2}\left(y^{\prime}{ }\right)}{\sqrt{\left(x-x^{\prime}\right)^{2}+\left(y-y^{\prime}\right)^{2}+\left(z-z^{\prime}\right)^{2}}}.
\end{align}
Substituting $x-x^{\prime} = x$,
\begin{equation}
    V_5(q) = \int_{-\infty}^{\infty} \frac{e^{i q x} \, d x}{\sqrt{x^{2}+\left(y-y^{\prime}\right)^{2}}} \int_{-\infty}^{\infty} d y \int_{-\infty}^{\infty} d y^{\prime}\,\xi_{0}^{2}\left(y\right)\xi_{0}^{2}\left(y^{\prime}\right).
\end{equation}
Using the integral identity for Bessel functions and Eq.\ (\ref{wf51}),
\begin{equation}\label{E5}
    V_5(q) = \frac{2}{\pi b^2} \int_{-\infty}^{\infty} d y \int_{-\infty}^{\infty} d y^{\prime} \, \exp \left({\frac{-y^2}{b^2}}\right) \exp \left({\frac{-y^{\prime 2}}{b^2}}\right) K_{0}\left(q\left|y-y^{\prime}\right|\right),
\end{equation}
where $K_{0}\left(q\left|y-y^{\prime}\right|\right)$ is the zeroth-order modified Bessel function of the second kind. Equation (\ref{E5}) can also be rewritten as
\begin{equation}\label{E6}
 V_5(q)= \frac{2}{\pi b^{2}} \int_{-\infty}^{\infty} d y \int_{-\infty}^{\infty} d y^{\prime} \, \exp{\left(-\frac{\left(y-y^{\prime}\right)^{2}}{2b^{2}}\right)} \exp{\left(-\frac{\left(y+y^{\prime}\right)^{2}}{2b^{2}}\right)} K_{0}\left(q\left|y-y^{\prime}\right|\right).
\end{equation}
Now transform $y$ and $y^{\prime}$ into center of mass coordinates. Given $\frac{y-y^{\prime}}{\sqrt{2}}=Y$ and $\frac{y+y^{\prime}}{\sqrt{2}}=Y^{\prime}$, the Jacobian transformation implies $d y \, d y^{\prime}=d Y \, d Y^{\prime}$ and so Eq.\ (\ref{E6}) becomes
\begin{equation}\label{E7}
 V_5(q)= \frac{2}{\pi b^{2}} \int_{0}^{\infty} \exp \left({\frac{-Y^{2}}{b^2}}\right) K_{0}(q \sqrt{2} Y) \, dY \int_{-\infty}^{\infty}  \exp \left({\frac{-Y^{\prime 2}}{b^2}}\right) \, d Y^{\prime}.
\end{equation}
Using the Gamma function and Bessel function integral identities, Eq.\ (\ref{E7}) simplifies to
\begin{equation}\label{analyticV5}
    V_5(q)  = \exp \left(\frac{q^{2} b^{2}}{4}\right) K_{0}\left(\frac{q^{2} b^{2}}{4}\right).
\end{equation}
The limiting behavior of harmonic-delta confinement is
\begin{align}
        V_5(q)& = -2\gamma +3\ln2 -2\ln(q\,b) \quad \text{for} \quad (q\,b) \rightarrow 0, \\
        & = \frac{\sqrt{2\pi}}{q\,b} \quad\quad\quad\quad\quad\quad\quad\quad\,\,\,\,\text{for}\quad (q\,b) \rightarrow \infty.\label{V5atlargeq}
\end{align}
%-------------------------------------------------------------------------------------------------

\section{Infinite square well confinement in the $y$ direction, delta function confinement in the $z$ direction}\label{AppendixF}
  Electrons are free to move in the $x$-direction and confined in the $yz$-plane using square well confinement in the $y$-direction and neglecting the motion of electrons in the $z$-direction, which is equivalent to the assumption of zero wire width. The one-electron wave function for this case is
\begin{equation}\label{twf6}
\psi(x, y, z)=\frac{e^{iqx}}{\sqrt{L}} \,\phi_{n}(y)\,\sqrt{\delta(z)}. 
\end{equation}
The ground state wave function for an infinite square well of width $b$ in the $y$-direction is
\begin{equation}\label{wf61}
   \phi(y)=\left(\frac{2}{b}\right)^{1 / 2} \sin \left(\frac{ \pi y}{b}\right), \quad 0 \leq y \leq b.
\end{equation}
Using Eqs.\ (\ref{twf6})  and (\ref{wf61}), the matrix elements of the Coulomb interaction are given by
\begin{eqnarray}
V_6(q)&=& \frac{4}{b^2} \int_{-\infty}^{\infty} dz^\prime\int_{-\infty}^{\infty}{\delta(z)\delta(z^\prime) \, dz} \int_{0}^{b} d y \int_{0}^{b} d y^{\prime} \, \sin^2\left(\frac{\pi y}{b}\right) \sin ^{2}\left(\frac{\pi y^{\prime}}{b}\right) \nonumber \\ &\times&  \int_{-\infty}^{\infty} \frac{e^{i q\left(x-x^{\prime}\right)} \, d\left(x-x^{\prime}\right)}{{\sqrt{\left(x-x^{\prime}\right)^{2}+\left(y-y^{\prime}\right)^{2}+\left(z-z^{\prime}\right)^{2}}}}.
\end{eqnarray}
Substituting $x-x^{\prime} = x$ and using the Dirac delta function's integral,
\begin{equation}\label{CIv6}
    V_6(q)= \frac{4}{b^2} \int_{0}^{b} d y \int_{0}^{b} d y^{\prime} \, \sin^2\left(\frac{\pi y}{b}\right) \sin ^{2}\left(\frac{\pi y^{\prime}}{b}\right) \int_{-\infty}^{\infty} \frac{e^{i qx} \, dx}{\sqrt{x^{2}+\left(y-y^{\prime}\right)^{2}}}.
\end{equation}
Using the integral identity for Bessel functions, Eq.\ (\ref{CIv6}) reduces to 
\begin{equation}
     V_6(q)= \frac{8}{b^2}   \int_{0}^{b} d y \int_{0}^{b} d y^{\prime} \, \sin^2\left(\frac{\pi y}{b}\right) \sin ^{2}\left(\frac{\pi y^{\prime}}{b}\right)K_{0}\left(q\left|y-y^{\prime}\right|\right),
\end{equation}
where $K_{0}\left(q\left|y-y^{\prime}\right|\right)$ is the zeroth-order modified Bessel function of the second kind.

%-------------------------------------------------------------------------------------------------
\section{}\label{AppendixG}
\noindent The self-energy contribution to the SSF [Eq.\ (\ref{S1self})] is
\begin{equation}
        S_1^\text{se}(q)= -\frac{1}{n\pi}\int^\infty_0 \chi_1^\text{se}(q,i\omega) \, d\omega,
\end{equation}
where, by Eq.\ (\ref{chi1self}), $\chi_1^\text{se}(q, i\omega)$ is proportional to a frequency-dependent part,
\begin{equation}
        \chi_1^\text{se}(q, i\omega) \propto \frac{\Omega_{k,q}^2 - \omega^2}{(\Omega_{k,q}^2 + \omega^2)^2}.
\end{equation}
To evaluate the integral
\begin{equation}
        \int^\infty_0 \chi_1^\text{se}(q,i\omega) \, d\omega
        = A \int_{0}^{\infty} \frac{\Omega^2 - \omega^2}{(\Omega^2 + \omega^2)^2}\, d\omega,   
\end{equation}
where $A$ is a proportionality constant,
we use the substitution $\omega = \Omega \tan \theta$.
The integral becomes
\begin{align}
        \int^\infty_0 \chi_1^\text{se}(q,i\omega) \, d\omega &= A\int_{0}^{\pi/2} \frac{\Omega^2 - (\Omega \tan \theta)^2}{[\Omega^2 + (\Omega \tan \theta)^2]^2} \left( \Omega \sec^2 \theta \,d\theta \right) \\
        &= \frac{A}{\Omega} \int_{0}^{\pi/2} \frac{1 - \tan^2 \theta}{\sec^2 \theta} d\theta = 0 .
\end{align}
Hence $S_1^\text{se}(q)=0$, i.e., the self-energy contribution to the SSF is zero.
\end{widetext}

%-------------------------------------------------------------------------------------------------
\bibliography{ref}

\end{document}